\documentclass[useAMS,usenatbib]{mnras}
\usepackage{natbib}
\usepackage[english]{babel}
\usepackage{graphicx,amsmath,amsfonts,amssymb}
\usepackage{colortbl}
\usepackage{tabu,booktabs}
\usepackage{xcolor}
\usepackage{xspace}
\usepackage{enumerate}
\usepackage{caption}
\usepackage{listings}
\usepackage{subfig}
\usepackage[justification=centering]{caption}

\usepackage[normalem]{ulem}

\def \etal {et~al.~}

\usepackage{gensymb,fontawesome}
\definecolor{linkcolor}{rgb}{0.1216,0.4667,0.7059}
\newcommand{\codeicon}{{\faCloudDownload}}
\newcommand{\codelink}{\href{https://github.com/TobiBu/ISM_subgrid_clumping/blob/main/ISM_subgrid_clumping.ipynb}{\faGithub}\,\,}
\newcommand{\plotlink}[1]{\href{https://github.com/TobiBu/ISM_subgrid_clumping/blob/main/plots/#1.pdf}{\codeicon}\,\,}
\newcommand{\oscaption}[2]{\caption{#2 \plotlink{#1}}}

\newcommand{\M}{\mathcal{M}}
\newcommand{\C}{\mathcal{C}}

\newcommand{\hMpc}{{\ifmmode{h^{-1}{\rm Mpc}}\else{$h^{-1}$Mpc}\fi}}
\newcommand{\Mpc}{{\ifmmode{{\rm Mpc}}\else{Mpc}\fi}}
\newcommand{\hkpc}{{\ifmmode{h^{-1}{\rm kpc}}\else{$h^{-1}$kpc}\fi}}
\newcommand{\kpc}{{\ifmmode{ {\rm kpc} }\else{{\rm kpc}}\fi}}
\newcommand{\kms}{{\ifmmode{ {\rm km\,s^{-1}} }\else{ ${\rm km\,s^{-1}}$ }\fi}}
\newcommand{\hMsun}{{\ifmmode{h^{-1}{\rm {M_{\odot}}}}\else{$h^{-1}{\rm{M_{\odot}}}$}\fi}}
\newcommand{\Msun}{{\ifmmode{{\rm M}_{\odot}}\else{${\rm M}_{\odot}$}\fi}}
\newcommand{\Mhalo}{{\ifmmode{M_{\rm halo}}\else{$M_{\rm halo}$}\fi}}
\newcommand{\Rvir}{{\ifmmode{R_{\rm vir}}\else{$R_{\rm vir}$}\fi}}
\newcommand{\Mstar}{{\ifmmode{M_{\rm star}}\else{$M_{\rm star}$}\fi}}
\newcommand{\Vrot}{{\ifmmode{V_{\rm rot}}\else{$V_{\rm rot}$}\fi}}
\newcommand{\ltsima}{$\; \buildrel < \over \sim \;$}
\newcommand{\gtsima}{$\; \buildrel > \over \sim \;$}
\newcommand{\lsim}{\lower.5ex\hbox{\ltsima}}
\newcommand{\gsim}{\lower.5ex\hbox{\gtsima}}

\def\lesssim{\mathrel{\hbox{\rlap{\hbox{\lower4pt\hbox{$\sim$}}}\hbox{$<$}}}}
\def\gtrsim{\mathrel{\hbox{\rlap{\hbox{\lower4pt\hbox{$\sim$}}}\hbox{$>$}}}}

\newcommand{\beq}{\begin{equation}}
\newcommand{\eeq}{\end{equation}}
\newcommand{\dps}{\displaystyle}

\def\beqa{\begin{eqnarray}}
\def\eeqa{\end{eqnarray}}
\def\LCDM{\ensuremath{\Lambda}CDM}

\def\head{ \vbox to 0pt{\vss \hbox to 0pt{\hskip 440pt\rm
      LA-UR-10-07069\hss} \vskip 25pt}}

\def \kms {\ifmmode  \,\rm km\,s^{-1} \else $\,\rm km\,s^{-1}  $ \fi }
\def \kpc {\ifmmode  {\rm kpc}  \else ${\rm  kpc}$ \fi  }  
\def \hkpc {\ifmmode  {h^{-1}\rm kpc}  \else ${h^{-1}\rm kpc}$ \fi  }  
\def \hMpc {\ifmmode  {h^{-1}\rm Mpc}  \else ${h^{-1}\rm Mpc}$ \fi  }  
\def \Mpch {\ifmmode  {h^{-1}\rm Mpc}  \else ${h^{-1}\rm Mpc}$ \fi  }  
\def \Msun {\ifmmode {\rm M}_{\odot} \else ${\rm M}_{\odot}$ \fi} 
\def \hMsun {\ifmmode h^{-1}\,\rm M_{\odot} \else $h^{-1}\,\rm M_{\odot}$ \fi}

\def \LCDM {\ifmmode \Lambda{\rm CDM} \else $\Lambda{\rm CDM}$ \fi}
\def \sig8 {\ifmmode \sigma_8 \else $\sigma_8$ \fi} 
\def \OmegaM {\ifmmode \Omega_{\rm m} \else $\Omega_{\rm m}$ \fi} 
\def \Omegab {\ifmmode \Omega_{\rm b} \else $\Omega_{\rm b}$ \fi} 
\def \OmegaL {\ifmmode \Omega_{\rm \Lambda} \else $\Omega_{\rm \Lambda}$\fi} 
\def \Deltavir {\ifmmode \Delta_{\rm vir} \else $\Delta_{\rm vir}$ \fi}
\def \rhocrit {\ifmmode \rho_{\rm crit} \else $\rho_{\rm crit}$ \fi}
\def \rhou {\ifmmode \rho_{\rm u} \else $\rho_{\rm u}$ \fi}
\def \zc {\ifmmode z_{\rm c} \else $z_{\rm c}$ \fi}

\def\araa{ARA\&A}%
\def\apj{ApJ}%
\def\apjl{ApJ}%
\def\apjs{ApJS}%
\def\aap{A\&A}%
\def\mnras{MNRAS}%
\def\ssr{Space~Sci.~Rev.}%
\def\head{ .ps \vbox to 0pt{\vss \hbox to 0pt{\hskip 440pt\rm
      LA-UR-10-07069\hss} \vskip 25pt}} 

\def \spose#1{\hbox  to 0pt{#1\hss}}  
\def \lta{\mathrel{\spose{\lower 3pt\hbox{$\sim$}}\raise 2.0pt\hbox{$<$}}}
\def \gta{\mathrel{\spose{\lower 3pt\hbox{$\sim$}}\raise 2.0pt\hbox{$>$}}}

\title[A statistical sub-grid model for the ISM]
{Escaping the maze: a statistical sub-grid model for cloud-scale density structures in the interstellar medium}

\author[Buck \etal] {
Tobias Buck$^{1}$\thanks{E-mail: tbuck@aip.de},
Christoph Pfrommer$^{1}$,
Philipp Girichidis$^{2,1}$,
Bogdan Corobean$^{1}$\\ 
$^1$Leibniz-Institut f\"ur Astrophysik Potsdam (AIP), An der Sternwarte 16, D-14482 Potsdam, Germany\\
$^2$Universität Heidelberg, Zentrum für Astronomie, Institut für Theoretische Astrophysik (ITA),\\ Albert-Ueberle-Str. 2, D-69120 Heidelberg, Germany\\
}

\begin{document}

\date{Accepted XXXX . Received XXXX; in original form XXXX}

\pagerange{\pageref{firstpage}--\pageref{lastpage}} \pubyear{2021}

\maketitle

\label{firstpage}


\begin{abstract}
The interstellar medium (ISM) is a turbulent, highly structured multi-phase medium. State-of-the-art cosmological simulations of the formation of galactic discs usually lack the resolution to accurately resolve those multi-phase structures. However, small-scale density structures play an important role in the life cycle of the ISM, and determine the fraction of cold, dense gas, the amount of star formation and the amount of radiation and momentum leakage from cloud-embedded sources. Here, we derive a \emph{statistical model} to calculate the unresolved small-scale ISM density structure from coarse-grained, volume-averaged quantities such as the \emph{gas clumping factor}, $\C$, and mean density $\left<\rho\right>_V$. Assuming that the large-scale ISM density is statistically isotropic, we derive a relation between the three-dimensional clumping factor, $\C_\rho$, and the clumping factor of the $4\pi$ column density distribution on the cloud surface, $\C_\Sigma$, and find $\C_\Sigma=\C_\rho^{2/3}$. Applying our model to calculate the covering fraction, i.e., the $4\pi$ sky distribution of optically thick sight-lines around sources inside interstellar gas clouds, we demonstrate that small-scale density structures lead to significant differences at fixed physical ISM density. Our model predicts that gas clumping increases the covering fraction by up to 30 per cent at low ISM densities compared to a uniform medium. On the other hand, at larger ISM densities, gas clumping suppresses the covering fraction and leads to increased scatter such that covering fractions can span a range from 20 to 100 per cent at fixed ISM density. All data and example code is publicly available at GitHub \codelink.
\end{abstract}

\noindent
\begin{keywords}
  methods: analytical -
  methods: statistical -
  methods: numerical -
  galaxies: formation - 
  galaxies: structure - 
  ISM: structure
  
 \end{keywords}

\vspace*{-1.5cm}
\section{Introduction} \label{sec:intro}

The interstellar medium (ISM) of galaxies is a highly structured, multi-component distribution of gas. The densest regions inside the ISM are giant molecular clouds (GMCs) with sizes of a few tens of parsecs \citep[e.g.][]{MivilleDeschenesMurrayLee2017}. Those regions of gas and dust undergo local gravitational collapse, forming dense cores in which new stars are born. The structure and morphology of GMCs is shaped by large-scale supersonic turbulence leading to the formation of density enhancements such as filaments, clumps, and cores \citep[e.g.][]{MacLowKlessen2004}.

A statistical representation of the influence of supersonic turbulence on the structure of molecular clouds is given by the probability distribution function (PDF) of the mass density. For isothermal, supersonic turbulent gas a lognormal density distribution is expected \citep[e.g.][]{vazquez1994,Nordlund1999,Ostriker2001,Klessen2000,Padoan2002,Krumholz2005,Wada2007,Hennebelle2008,Federrath2010,Konstandin2012}. Recent observations by \citet{kainulainen2009} showed further that also the column density PDFs of GMCs exhibit a lognormal distribution.

In general, star formation is expected to occur in the coldest and densest parts of the ISM dominated by molecular gas \citep[e.g.][]{Kennicutt2012,Girichidis2020}. 
Successive mechanical and radiative energy input from young massive stars will disperse and ionize the dense gas damping further star formation, although the effect of self-shielding can diminish radiation feedback, and thus may enable further star formation. In fact, the existence of H$\alpha$ emission around young star forming regions is a prominent signature of rapidly operating feedback processes in and around those star-forming regions \citep[e.g][]{Kreckel2018}.

Accurately modeling the star formation–feedback cycle is a key prerequisite in shaping galactic properties both on the scales of the ISM \citep[e.g.][]{Walch_2015,Girichidis2016,Semenov2017,Semenov2021,KimEtAl2020,Gutcke2021,RathjenEtAl2021} as well as the formation of galaxies in the cosmological context \citep[e.g.][]{Brook2012,Agertz2013,Agertz2015,Agertz2016,Wang2015,Grand2017,Hopkins2018,Buck2017,Buck2020,Applebaum2020,Applebaum2021}. In the past few years important progress has been made in modeling star formation, feedback \citep[e.g.][]{Marinacci2019,Emerick2019,Benincasa2020,Smith2020}, radiation hydrodynamics \citep[e.g.][]{Rosdahl2015,Kannan2014,Kannan2016,Kannan2020,Emerick2018,Obreja2019}, non-thermal feedback processes \citep[e.g.][]{GirichidisEtAl2016a,Pfrommer2017,Butsky2020,Buck2020a} as well as the chemistry of the ISM \citep[e.g.][]{Robertson2008,Gnedin2010,Gnedin2011, Hopkins2011,Christensen2012,Buck2021}.
Despite the great advancements in numerical resolution and the successes of the models, 
there is still great uncertainty in the relevant physical processes and their specific numerical implementation as subgrid recipes \citep[see e.g.][for recent reviews]{Somerville2015,Naab2017,Vogelsberger2020}. 

Implementing physical processes below the resolution limit of the simulation poses substantial challenges for current galaxy formation models \citep[e.g.][]{Keller2019,Munshi2019,Genel2019,Buck2019,Dutton2020}. Depending on the implementation of the subgrid models either careful calibration against observations or fine tuning of parameter combinations are required. Often, this procedure results in somewhat disconnecting the subgrid model from the resolved scales of the simulation in the sense that (i) the calibration is not obtained from coarse-graining small-scale simulations and (ii) only few models do not require re-tuning parameters if run at different numerical resolution.

A promising avenue for improving the current state-of-the-art is to develop new subgrid-scale models which derive statistical properties of the gas from high-resolution ISM simulations and apply those to large-scale (cosmological) simulations in which those scales are not directly resolved, thus establishing a connection with the resolved scales. Here we develop such a statistical model for the density distribution or porosity of the ISM on scales of a few ten parsecs. The logic is the following: In coarse grained galaxy formation simulation each resolution element carries resolved information such as its (volume) averaged density, $\left<\rho\right>_V^R$ and its spatial size, $R$, while its substructure is a priori unknown and depends on unresolved physics. On the other hand, high resolution simulations of either single GMCs or whole patches of galactic discs are able to follow the physics on much smaller scales. Coarse graining their results on the resolution scale of cosmological simulations then allows to use these high-resolution simulations as super-resolution models. Here we characterize the density substructure via the density distribution function and tie its characteristic parameters, i.e., its width and peak position, to fundamental parameters of coarse resolution elements such as their volume averaged density, $\left<\rho\right>_V^R$ and their spatial size, $R$. This will allow to estimate the density sub-structure of coarse resolution elements simply from those properties. The only free parameter of our model is the clumping factor of the density whose dependence on spatial scale, $R$, and average density, $\left<\rho\right>_V^R$, can be robustly derived from high-resolution simulations. When applying the model, the appropriate values for this parameter can then be statistically sampled from the derived distributions. 

In this work we set out to derive a parametrization of the density sub-structure on scales that can be resolved in current cosmological simulations. We are especially interested in the three-dimensional density structure as seen from the position of potential sources such as stars and its connection to the $4\pi$ column density distribution around those sources on the surface of spheres of radius $R$.

Our newly derived model is ideally suited to estimate the fraction of dense gas below the resolution limit \citep[see Section 2.2 of][]{Dominguez2014} or to calculate H$_2$ fractions \citep[e.g.,][]{Gnedin2009,Christensen2012} or to calculate the surface mass density distribution of GMCs from their average density alone. The model is able to self-consistently predict the $4\pi$ column density distribution around stars and the corresponding covering fraction (the number of optically thick sight-lines) of ISM clouds. It is therefore well suited to be coupled with e.g. radiation-hydrodynamic schemes to calculate the (UV) photon escape fractions for radiation sources from e.g. their birth clouds \citep[see also][for a similar approach to subgrid modelling IGM clumping]{Mao2020,Bianco2021}. 

The remaining paper is structured as follows. In Section~\ref{sec:theo} we discuss the PDF of the ISM and explore in particular the log-normal density distribution and its statistics. In Section~\ref{sec:results} we use this formalism to develop a statistical sub-grid model for the density distribution of the ISM. We use this model to derive the cloud scale column density distribution and calculate the $4\pi$ covering fraction of gas clouds as a function of their average density in turbulent box simulations. We further validate and compare our model to high-resolution magneto-hydrodynamic simulations of \citet{GirichidisEtAl2018b} from the SILCC project \citep{Walch_2015, Girichidis2016}. In Section~\ref{sec:appl} we discuss three potential applications of the model in coarse-grained simulations to (i) calculate the dense gas fraction of resolution elements and thus estimate star formation efficiencies, (ii) model energetic and radiative feedback efficiencies, and (iii) estimate radiation leakage from gas cloud embedded sources. We end this paper in Section~\ref{sec:dis} with a summary and conclusions.

\section{The Probability Distribution Function of the ISM} \label{sec:theo}

In this chapter we will derive a connection between the gas density clumping factor and the variance of the gas density PDF and connect those properties to coarse-grained quantities such as the average density on a given spatial scale $R$. For this, we define the {\em volume-weighted} gas density PDF, $f_{V}(\rho)$, which describes the fractional volume per unit density ($V^{-1}\rmn{d}V/\rmn{d}\rho$) density PDF, $f_{M}(\rho)$, which describes the fractional mass per unit density ($M^{-1}\rmn{d}M/\rm{d}\rho$). We can relate $f_{V}(\rho)$ to $f_{M}(\rho)$, using the fact that $f_{M}\propto\rmn{d}M/\rm{d}\rho$ and $f_{V}\propto\rmn{d}V/\rm{d}\rho$. From this we have 
\begin{align}
f_{M}(\rho) \propto\frac{\rmn{d}M}{\rmn{d}V}\frac{\rmn{d}V}{\rmn{d}\rho}\propto\rho\,f_{V}(\rho).
\end{align}

\subsection{Log-normal density PDFs}
\label{sec:lognorm}

The density PDF in the ISM varies between the different regimes and spatial scales. In low-density regions, in which self-gravity is negligible and turbulence dominates, the density PDF can be approximated by a log-normal distribution \citep[e.g.][]{Ostriker2001}. In dense regions that are gravitationally collapsing, the PDF develops a high-density power-law tail \citep[e.g.][]{Klessen2000, Slyz2005, Federrath_Klessen_2012, Girichidis2014}. On scales of GMCs and above, the density PDF $f(\rho)$ is in agreement with a single log-normal probability distribution function \citep[LN-PDF e.g.][]{Berkhuijsen2008,Berkhuijsen_2015}:
\begin{align}
f_{V}(\rho)\mathrm{d} \rho = \frac{1}{\sqrt{2\pi}\sigma} \exp{\left[-\frac{(\ln \rho- \mu_\rho)^2}{2\sigma_\rho^2}\right] } \frac{\mathrm{d} \rho}{\rho},
\label{eq:lognorm}
\end{align}
where $\mu_\rho=\ln\rho_0$, $\rho_0$ is the median of the distribution and $\sigma_\rho$ is the width of that distribution\footnote{Note that the LN-PDF is in other numerical studies sometimes also written as a function of $s\equiv\ln(\rho/\rho_0)$ normalizing the density to the average density $\rho_0=M/L^3$ in the simulation domain. Here we work with non-normalised quantities but converting between the two approaches can easily be achieved by a coordinate transformation.}.
Noting that $\rho>0$, the corresponding cumulative function for the log-normal density distribution is given by:
\begin{align}
P\left(\rho\leq\rho_\rmn{thr}\right) &= \frac{1}{2}\left(1+\mathrm{erf\left[\frac{\ln\left(\rho_\rmn{thr}/\rho_0\right)}{\sqrt{2}\sigma_\rho}\right]}\right).
\label{eq:cumulative}
\end{align}

From Eq.~\eqref{eq:lognorm} we can derive the volume-averaged density $\langle \rho\rangle_V$ of the gas following a log-normal distribution:
\begin{align}
\langle\rho\rangle_{V} &= \int^\infty_{0} \rho f_{V}(\rho) \mathrm{d} \rho= \rho_0 \, \exp\left(\frac{\sigma_\rho^2}{2}\right).
\label{eq:vavr}
\end{align}

Correspondingly, the mass-weighted density is given by
\begin{align}
\langle\rho\rangle_M  &= \int^\infty_{0} \rho f_{M}(\rho) \mathrm{d} \rho = \int^\infty_{0} C_0 \rho^2 f_{V}(\rho) \mathrm{d} \rho,
\label{eq:massavr}
\end{align}
where we have used the fact that $f_{M}(\rho)=C_0\rho\,f_{V}(\rho)$.
The constant $C_0$ can easily be determined by the requirement that the PDF is normalised, i.e. $\int^\infty_{0} f_{M}(\rho) \mathrm{d} \rho = \int^\infty_{0} C_0 \rho f_{V}(\rho) \mathrm{d} \rho = 1$, which gives 
\begin{align}
C_0^{-1} = \rho_0 \, \exp\left(\frac{\sigma_\rho^2}{2}\right)
\end{align}
using Eq.~\eqref{eq:vavr}. Thus, the mass-weighted average density evaluates to \citep[see also][]{Li2003}
\begin{align}
\langle\rho\rangle_M &= \rho_{0} \exp\left(\frac{3\sigma_\rho^2}{2}\right).
\label{eq:massavr}
\end{align}

These relations define a simple form for the dispersion $\sigma$ of the LN-PDF: 
\begin{align}
\sigma_\rho^2 = \ln \left( \frac{\langle\rho\rangle_M}{\langle\rho\rangle_V} \right),
\label{eq:sigma}
\end{align}
which we can also rewrite using $\rho_0$, 
\begin{align}
\label{eq:sigma1}
\sigma_\rho^2 = 2 \ln \left(\frac{\langle\rho\rangle_V}{\rho_0} \right) 
= \frac{2}{3} \ln \left(\frac{\langle\rho\rangle_M}{\rho_0}\right). 
\end{align}
Under the assumption of nearly constant characteristic density $\rho_0$, Eq.~\eqref{eq:sigma1} implies that the dispersion $\sigma_\rho$ is proportional to the total mass of the system.

The above equations show that, for a stable, uniform system, i.e. $\left<{\rho}\right>_V = \rho_0$, $\sigma_\rho$ will be zero. 
On the other hand, if $\left<{\rho}\right>_V \rightarrow \infty$, $\sigma_\rho \rightarrow \infty$, which in fact resembles a dynamically unstable system.
Therefore, we expect that in a globally stable, inhomogeneous system $\sigma_\rho$ will take on numbers in an appropriate range.

In order to establish a connection between $\left<\rho\right>_V$ and $\left<\rho\right>_M$, we show that the clumping factor for the density $\C_\rho$ is related to their ratio:
\begin{align}
\C_\rho &\equiv \frac{\left<\rho^2\right>_V}{\left<\rho\right>_V^2}
= \frac{\left<\rho\right>_M}{\left<\rho\right>_V}=\exp\left({\sigma_\rho^2}\right).
\label{eq:clump}
\end{align}
Combining this definition with Eq.~\eqref{eq:sigma} and the fact that the median of the log-normal is given by the median of the density distribution $\rho_0$, Eq.~\eqref{eq:sigma1} relates $\sigma_\rho^2$ to the ratio of volume weighted mean over median. 

Numerical studies of ISM turbulence \citep[e.g.][]{Federrath2010} have further established a relation between clumping factor, $\C_\rho$, with the turbulence parameter $b$ and the Mach number, $\M$:
\begin{align}
\C_\rho = 1 + b^2\M^2.
\label{eq:mach}
\end{align}
This can be further related to the column density fluctuations under specific assumptions \citep[see e.g.][]{Burkhart2012}.

\subsection{Cloud-scale column density distribution} \label{sec:method}

It is well established that inter-stellar gas clouds are not entities of uniform density but highly structured objects \citep[e.g.][]{HeyerDame2015} with dense clumps and filaments embedded into lower density, hot gas. In this work, we refer to the column density $\Sigma=\rmn{d}M/\rmn{d}A$ as the projected density onto the surface of a spherical gas cloud with radius $R$, total mass $M$, and $\rmn{d}M$ denoting the cone mass subtended by the area element $\rmn{d}A$. The spherical area element is here given by $\rmn{d}A=(r\rmn{d}\theta)\,(r\sin{\theta}\,\rmn{d}\phi)$. Later in Section \ref{sec:data} we use the HEALpix formalism \citep{Gorski_2005} to calculate column density distribution on the surface of simulated spheres.

Thus, the column density at the surface of the sphere inherits this property of the ISM and will be highly structured as well.
That means in a structured medium there will be a difference between the surface mass density measured over an area $A$ (like e.g. the whole surface of a sphere) which is equal to the \textit{area-weighted surface mass density},
\begin{align}
\left<\Sigma\right>_A&\equiv\frac{\int_A\Sigma\,\mathrm{d}A}{\int_A\mathrm{d}A},
\label{eq:Asurf}
\end{align}
and the surface mass density at which most of the mass is found, the \textit{mass-weighted surface mass density} $\left<\Sigma\right>_M$.
\begin{align}
\left<\Sigma\right>_M
&\equiv\frac{\int\Sigma\,\mathrm{d}M}{\int\,\mathrm{d}M}
=\frac{\int_A\Sigma^2\,\mathrm{d}A}{\int_A\Sigma\,\mathrm{d}A}\,. 
\end{align}
For a uniform medium the two quantities $\left<\Sigma\right>_M$ and $\left<\Sigma\right>_A$  will be the same. In a highly clumped medium where most of the mass resides in small, high column density regions spread over large, low column density areas the two quantities will significantly differ. 
In order to establish a connection between those two quantities, we show that the clumping factor for the column density $\C_\Sigma$ is related to the ratio of $\left<\Sigma\right>_M$ and $\left<\Sigma\right>_A$:
\begin{align}
\C_\Sigma &\equiv \frac{\left<\Sigma^2\right>_A}{\left<\Sigma\right>_A^2}
= \frac{\left<\Sigma\right>_M}{\left<\Sigma\right>_A},
\label{eq:clumping}
\end{align}
\citep[see also Eq. 3 in][]{Leroy2013}. This definition is equivalent to the definition of the clumping factor via the volume density $\rho$ as given in Eq.~\eqref{eq:clump}. 

\begin{figure*}
  \centering
  \includegraphics[width=.49\textwidth]{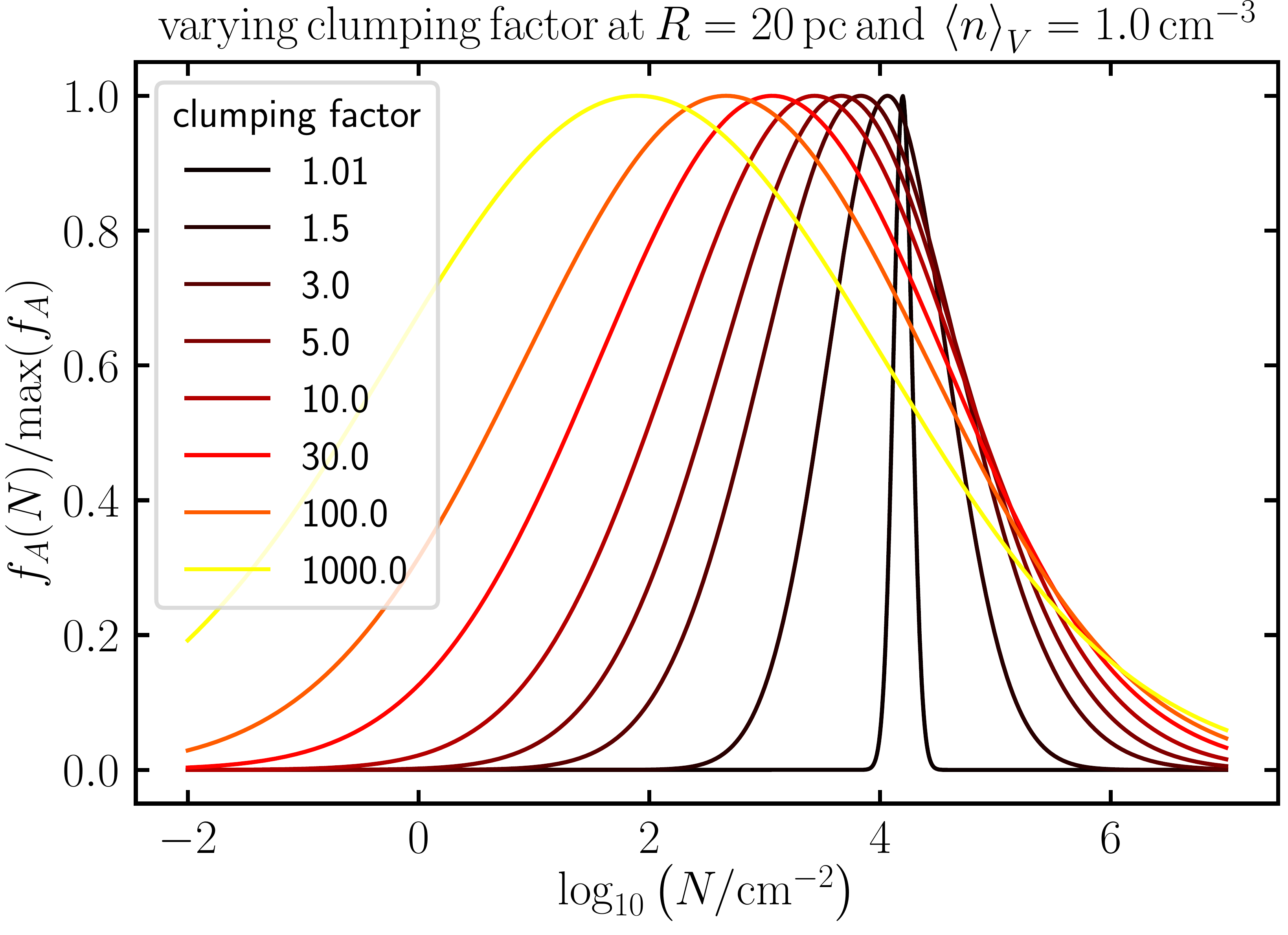}
  \includegraphics[width=.49\textwidth]{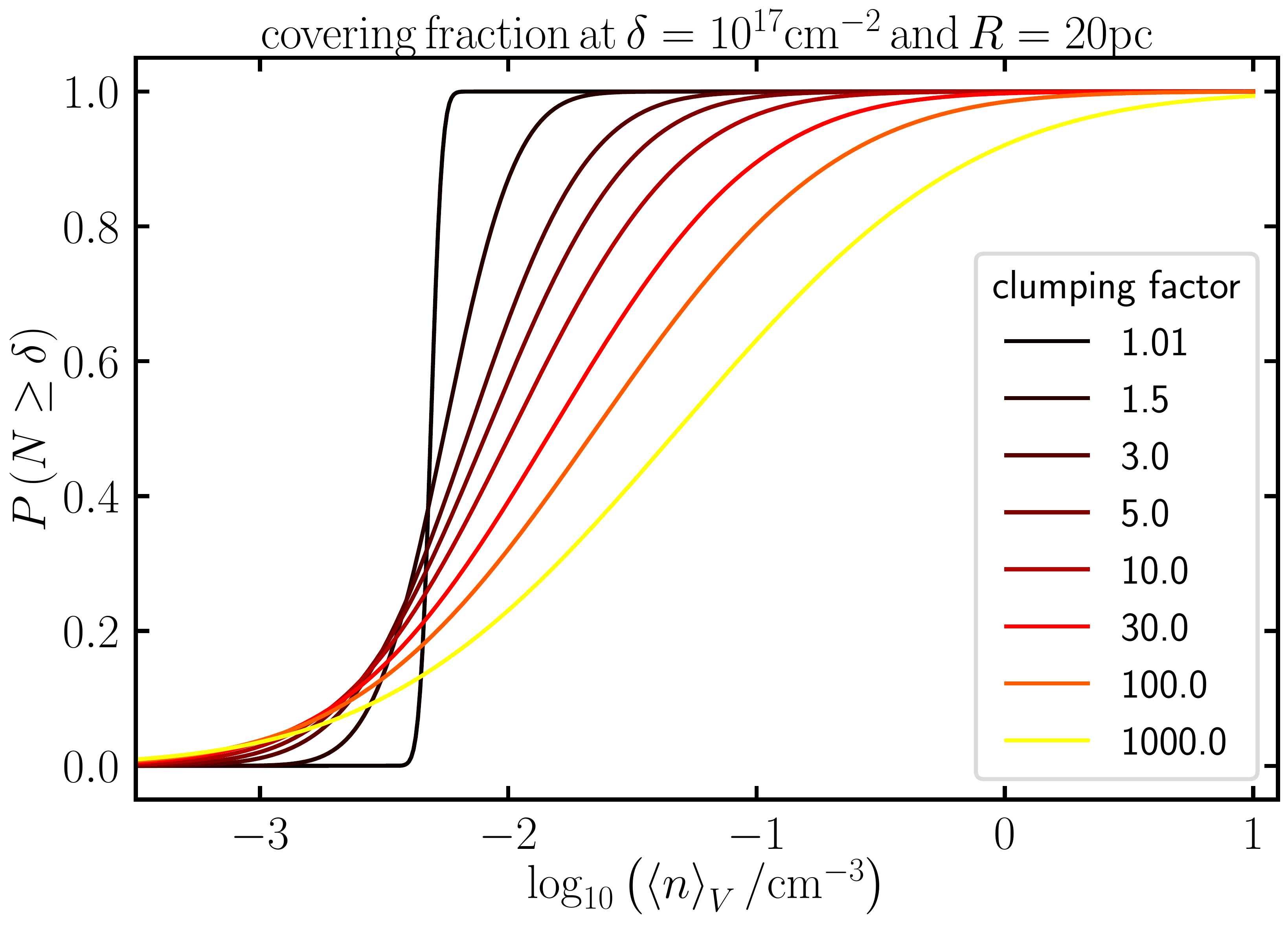}
  \oscaption{cov_frac}{The left-hand panel shows the column density distribution of a representative spherical patch of the ISM based on our analytical model (Eqs.~\ref{eq:sigma_lognorm} and \ref{eq:clump_sigma}). To exemplify the impact of ISM clumping on the width of the column density distribution we show results for different clumping factors. Note, we renormalise those curves to emphasize the change in peak position and width. In the right-hand panel we show the covering fraction as a function of ISM density as derived in Eq.~\eqref{eq:err3} (see Section \ref{sec:data} for more details). For this plot we have chosen a spherical region of 20 pc radius and a column density threshold of $10^{17}$ cm$^{-2}$ chosen to roughly match the Lyman limit HI column density which is given by $N_{\rm{HI}^{\rm{thick}}}\sim7.3\times10^{17}$~cm$^{-2}$. Again we exemplify the impact of ISM clumping by showing the results for different clumping factors.}
  \label{fig:theory}
\end{figure*}

For a uniform density distribution where by definition $\C_\rho=1$ we expect the surface mass density distribution to be a delta distribution around the average area weighted surface mass density. The more clumpy the medium, the broader the distribution becomes because we will find some areas on the sphere that are over-dense due to clumping and others, which (for mass conservation reasons) are under-dense.

We can further derive a link between the volume-weighted density and the area-weighted surface density as follows:
\begin{align}
\left<\rho\right>_V = \frac{\int_V\rho\,\mathrm{d}V}{\int_V\mathrm{d}V} =\frac{\int_A\Sigma\,\mathrm{d}A}{\int_V\mathrm{d}V}\frac{\int_A\mathrm{d}A}{\int_A\mathrm{d}A} =\left<\Sigma\right>_A\frac{3}{R},
\label{eq:relation}
\end{align}
where we have used the definition of the area-weighted surface density from Eq.~\eqref{eq:Asurf} and the relation of volume and surface area of a sphere, $V/A=R/3$, where $A=4\pi R^2$. 

We adopt the plausible assumption that the density distribution in the ISM is isotropic and homogeneous if averaged over sufficiently large regions  such that large-scale density gradients or edge effects can be neglected. This means that the density distribution along the three spacial dimensions are independent\footnote{This is of course a simplification and self-gravity or convergent flows might break this independence and correlate the density distribution along different axes at some spatial scale.} and also log-normally distributed.

Therefore, it follows that the column density $\Sigma$ on the sphere surface (the two dimensional density distribution), obtained by projecting the three dimensional density distribution $f_V$ along the radial axis is also log-normally distributed. The area-weighted LN-PDF of $\Sigma$ then reads:
\begin{align}
f_{A}(\Sigma)\mathrm{d} \Sigma = \frac{1}{\sqrt{2\pi}\sigma_\Sigma} \exp{\left[-\frac{(\ln \Sigma- \mu_\Sigma)^2}{2\sigma_\Sigma^2}\right] } \frac{\mathrm{d} \Sigma}{\Sigma},
\label{eq:sigma_lognorm}
\end{align}
where $\mu_\Sigma=\ln\Sigma_0$ is the characteristic column density, $\sigma_\Sigma$ is the width of that distribution, and $\Sigma_0=M/(4\pi R^2)$.

Following the derivation in Section~\ref{sec:lognorm} and replacing $V$ by $A$ and $\rho$ by $\Sigma$, we find the area- and mass-weighted column densities:
\begin{align}
\langle\Sigma\rangle_{A} &= \Sigma_0 \exp\left(\frac{\sigma_{\Sigma}^2}{2}\right),
\label{eq:surfsigma}\\
\langle\Sigma\rangle_{M} &= \Sigma_0 \exp\left(\frac{3\sigma_{\Sigma}^2}{2}\right).
\end{align}
This enables us to relate the clumping factor to the width $\sigma_{\Sigma}$ via \citep[see also][]{Gnedin2009,Lupi2018}:
\begin{align}
\ln \C_\Sigma &= \ln\left(\frac{\langle\Sigma\rangle_{M} }{\langle\Sigma\rangle_{A}}\right) = \sigma_{\Sigma}^2.   
\label{eq:clump_sigma}
\end{align}
Combining Eq.~\eqref{eq:surfsigma} with Eq.~\eqref{eq:relation} and replacing $\sigma_{\Sigma}^2$ by Eq.~\eqref{eq:clump_sigma} we can express $\Sigma_0$ as:
\begin{align}
\Sigma_0 &= \frac{R}{3}\left<\rho\right>_V\exp\left(-\frac{\ln \C_\Sigma}{2}\right)
= \frac{R}{3}\frac{\left<\rho\right>_V}{\C_\Sigma^{1/2}}.
\label{eq:sigma0}
\end{align}
Equation \eqref{eq:sigma0} expresses $\Sigma_0$ as a function  of volumetric and projected properties of the gas cloud. 

To obtain an expression relating the two- and three-dimensional clumping factors, $\C_\Sigma$ and $\C_\rho$, we start with Eq.~\eqref{eq:vavr},
\begin{align}
    \left<\rho\right>_V=\frac{M}{V}=\frac{M}{\frac{4\pi}{3} R^3} = \rho_0 \, \exp\left(\frac{\sigma_\rho^2}{2}\right). \label{eq:vavr2}
\end{align}
Using the definition of the surface density, we obtain
\begin{align}
    \left<\Sigma\right>_A = \frac{M}{A}=\frac{M}{4\pi R^2} = \Sigma_0 \exp\left(\frac{2}{3}\frac{\sigma_\rho^2}{2}\right),
    \label{eq:derivation}
\end{align}
where we expressed the radius $R$ in the last step in terms of $\left<\rho\right>_V$ via Eq.~\eqref{eq:vavr2}.
Comparing Eqs.~\eqref{eq:derivation}  and \eqref{eq:surfsigma} shows that $\sigma_{\Sigma}^2 = \frac{2}{3} \sigma_\rho^2$. This translates into a relation between the two- and three-dimensional clumping factors, $\C_\Sigma$ and $\C_\rho$, via Eqs.~\eqref{eq:clump} and \eqref{eq:clump_sigma}, resulting in:
\begin{align}
    \ln\C_\Sigma=\frac{2}{3}\ln\C_\rho.
    \label{eq:C_rho_vs_C_Sigma}
\end{align}

In Appendix \ref{app:c_vs_c}, in Fig.~\ref{fig:c_vs_c} we show the empirical relation between $\C_\rho$ and $\C_\Sigma$ as quantified from turbulent box simulations and multi-physics simulations from the SILCC simulation project \citep[][see Sections~\ref{sec:turb} and \ref{sec:sim} for more details]{GirichidisEtAl2018b}. This figure shows that Eq.~\eqref{eq:C_rho_vs_C_Sigma} is indeed on average valid, independent of the size of the sphere. However, we find some scatter around the main relation that is growing as a function $\C_\rho$.
We will use Eq.~\eqref{eq:C_rho_vs_C_Sigma} to re-write Eq.~\eqref{eq:sigma0}:
\begin{align}
\Sigma_0 &= \frac{R}{3}\left<\rho\right>_V\C_\rho^{-1/3}.
\label{eq:sigma0_1}
\end{align}

\subsection{Covering fractions from column density distributions}

For many astrophysical phenomena (especially the ones including sources and sinks) one of the most important quantities is the distribution of column density values. Especially for processes including radiation the interesting property is the fraction of lines-of-sight, $N\left(\geq\delta\right)$, above a given threshold of mass surface density, $\delta$, above which the radiation is completely absorbed.

We find that the fraction of sight lines with column density values above a given threshold $\log\delta$, is mathematically given by the cumulative function of the log-normal distribution which is known as the error function \citep[see also][for a similar argument]{Elmegreen2002,Wada2007}. In fact, the fraction of regions with column density larger than a threshold $\ln\delta$ is given by:
\begin{align}
P(\Sigma\geq\delta)=\frac{1}{2}\left(1 - \mathrm{erf\left[\frac{\ln\left(\delta/\Sigma_0\right)}{\sqrt{2}\sigma_{\Sigma}}\right]}\right)
\label{eq:err}
\end{align}
for column densities following a log-normal distribution with characteristic surface density $\Sigma_0$ and width $\sigma_{\Sigma}$. In the following, we refer to $P(\Sigma\geq\delta)$ as the covering fraction.

We are now fully equipped to derive a model for the column density distribution on the sphere which is solely dependent on the mean density inside the sphere, $\left<\rho\right>_V$, and the clumpiness of the medium given by the value of $\C_\rho$. As such we are able to calculate the fraction of the $4\pi$ sphere around a point $x$ in the ISM that is covered as a function of the average density inside the sphere of radius $R$.
Combining Eqs.~\eqref{eq:surfsigma} and \eqref{eq:sigma0_1} and inserting them into eq.~\eqref{eq:err} for the covering fraction we arrive at our final equation:
\begin{align}
&P(\Sigma\geq\delta) = f(\left<\rho\right>_V\vert R,\C_\rho,\delta)\\ 
&= \frac{1}{2}\left(1 - \mathrm{erf}\left[\frac{\ln\left(\frac{\dps3\delta}{\dps R\left<\rho\right>_V}\C_\rho^{1/3}\right)}{\sqrt{2\ln \C_\Sigma}}\right]\right)\\
&= \frac{1}{2}\left(1 - \mathrm{erf}\left[\frac{\ln\hat{\delta} - \hat{\mu}}{\sqrt{2}\hat{\sigma}}\right]\right).
\label{eq:err3}
\end{align}
In the last step we have introduced a scaled surface density threshold, $\hat{\delta}$, a scaled peak position, $\hat{\mu}$, and a scaled width, $\hat{\sigma}$, which are defined as follows: 
\begin{align}
    \hat{\delta}\equiv\frac{3\delta}{R\left<\rho\right>_V}, \qquad
    \hat{\mu}\equiv-\frac{1}{3}\ln\C_\rho, \qquad
    \hat{\sigma}\equiv\sqrt{\frac{2}{3}\ln \C_\rho}.
\end{align}

Figure~\ref{fig:theory} shows how the theoretical column density distribution and correspondingly the covering fraction of the gas cloud varies as a function of clumping factor. The left panel shows how the peak and width of the LN-PDF change as a function of clumping factor $\C_\rho$ and the right panel shows how a clumpy medium enhances the covering fraction for low densities clouds (below a characteristic density equivalent to to the threshold column density $\delta$) and reduces the covering fraction for high density clouds at a fixed mean density. To create these curves, in the left panel we plot Eq.~\eqref{eq:sigma_lognorm} replacing $\Sigma_0$ in $\mu_\Sigma=\ln\Sigma_0$ with Eq.~\eqref{eq:sigma0_1} and in the right panel we plot Eq~\eqref{eq:err}.

\section{A statistical model for sub-grid ISM clumping and $4\pi$ cloud-scale covering fractions} \label{sec:results}

The aim of this work is to derive a statistical model for the sub-grid structure of the ISM which can readily be applied to simulations with a coarser resolution such as cosmological simulations of the formation of Milky Way-like galaxies. Those simulations usually lack the spatial resolution to properly resolve the multi-phase nature of the ISM. In Section~\ref{sec:theo} we have derived our theoretical model for the sub-grid clumping of the ISM and its effect on e.g. the covering fraction of (molecular) gas clouds. In order to gauge the performance of our model we first compare it to idealized models of driven turbulence before applying it to models of the ISM. For the former step we use the simulations of driven isothermal turbulence presented in \citep{KonstandinEtAl2015,KonstandinEtAl2016}, which are performed using the \textsc{Flash} code \citep{Fryxell2000,Dubey2008} in a periodic box with a uniform resolution of $1024^3$ cells. The forcing follows \citet{SchmidtEtAl2009} with a natural mix of solenoidal and compressive driving on scales of $n=1-3$, where $nk=2\pi/L_\mathrm{box}$. The second set of models uses state-of-the-art resolved ISM simulations from the SILCC project \citep{Walch_2015, Girichidis2016}. For detailed descriptions of the numerical setup of the simulations and the physics models employed we refer the reader to these references. The ISM simulations used here are the higher resolution runs described in detail in \citet{GirichidisEtAl2018b}. For completeness, below we briefly describe the simulation and our procedure to extract cloud-scale data from it to compare to our model.

\subsection{Data extraction using \texttt{HEALPix}}
\label{sec:data}

We statistically analyse different regions in the simulation domain by choosing randomly placed positions, $\mathbfit{x}$, and investigating spheres around that position with a fixed given radius, $R$. For each sphere we compute the average density as the main quantity of the analysis volume. To characterize the distribution of $4\pi$ column densities that are computed by projecting the density within cones from the centres of the spheres to the radius, we make use of the \texttt{HEALPix}\footnote{\url{http://healpix.sf.net}} \cite[]{Gorski_2005} tessellation of the unit sphere. \texttt{HEALPix} divides the surface of the unit sphere into $N_{\rmn{pix}}$ quadrilateral, curvilinear pixels of varying shapes but equal area. The resolution depends on the $N_\rmn{side}$ parameter, i.e. $N_{\rmn{pix}} = 12 \times N_{\rmn{side}}^2$, where $N_{\rmn{side}}$ must be a power of two. For our analysis here we have chosen $N_{\rmn{side}} = 4$ corresponding to 192 cells on the sphere surface. Tests with larger values for $N_{\rmn{side}}$ show that our results are robust against changes in the \texttt{HEALPix} resolution.

We use the python implementation of the \texttt{HEALPix} algorithm from the \texttt{healpy} package \citep{healpy} in order to project the mass inside a spherical region of radius $R$ onto its surface and evaluate the resulting distribution of column density pixel values. From the \texttt{HEALPix} tesselation we can further directly derive the surface density clumping factor, $\C_\Sigma$ as defined in Eq.~\eqref{eq:clumping} and the covering fraction by counting the number of pixels with column density \textit{above} a certain threshold value, $\delta$, divided by the total number of pixels (in our case 192 pixels).

\subsection{Driven turbulence simulations}
\label{sec:turb}

\begin{figure*}
  \centering
  \includegraphics[width=.33\textwidth]{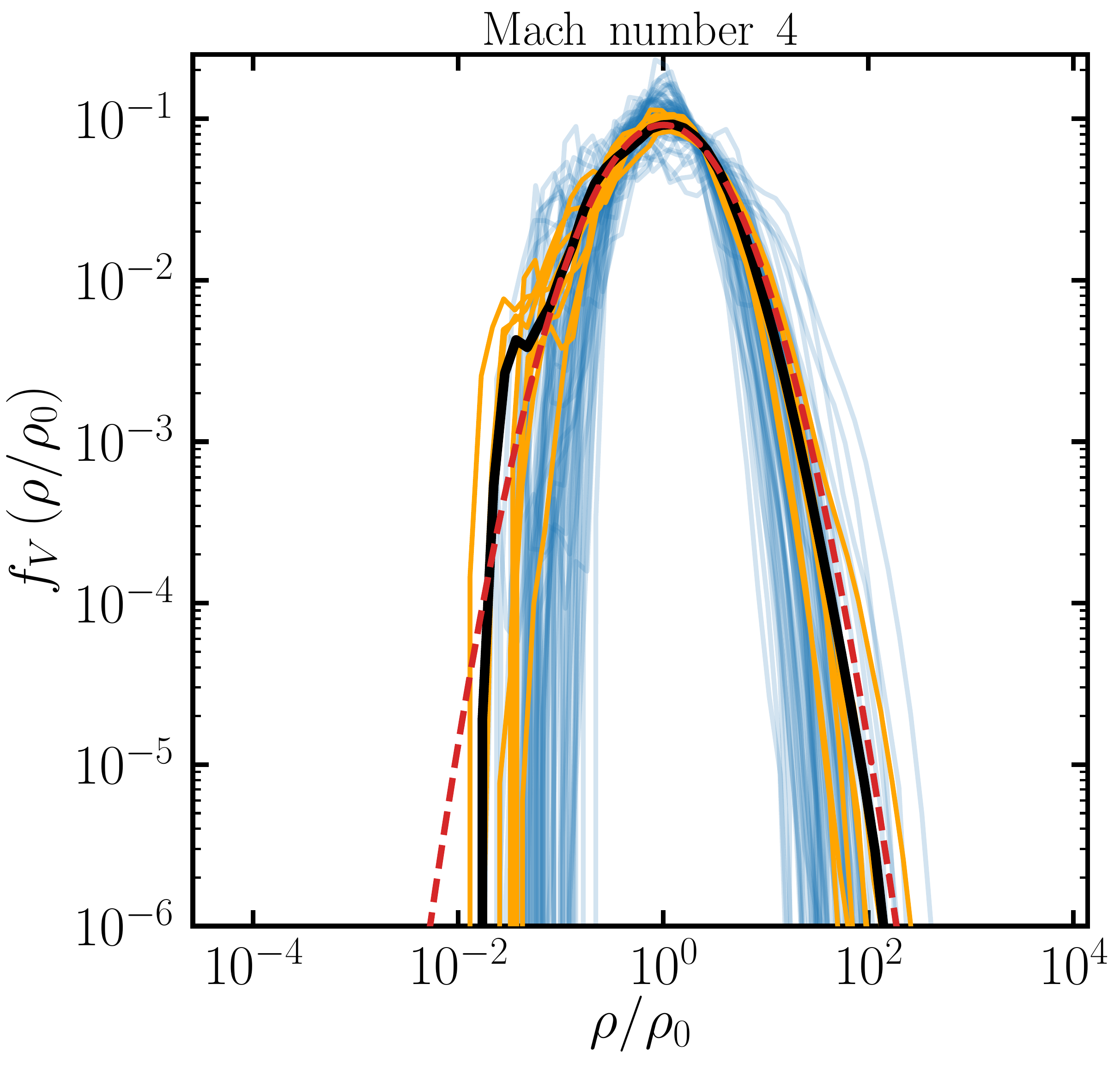}
  \includegraphics[width=.33\textwidth]{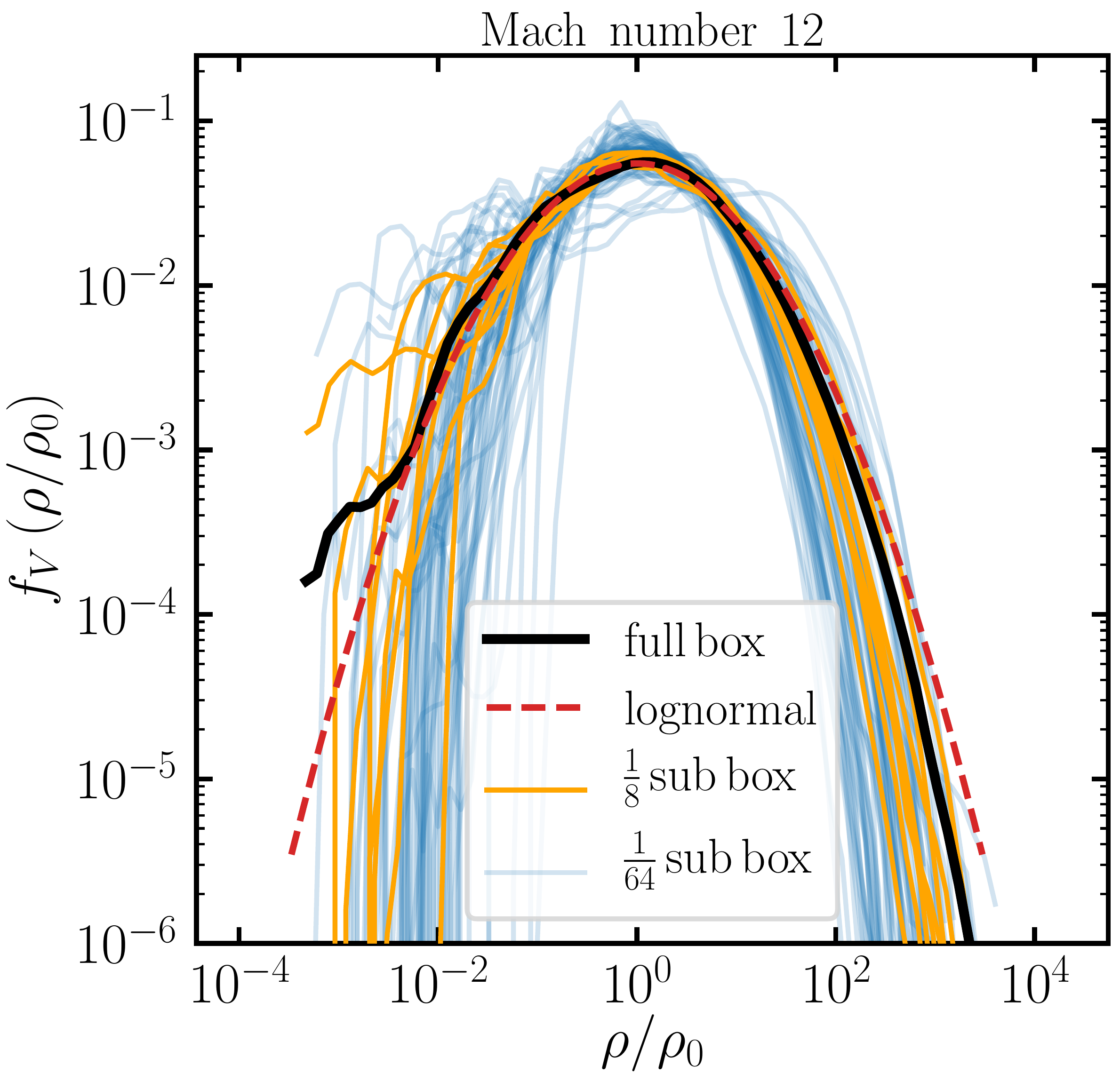}
  \includegraphics[width=.33\textwidth]{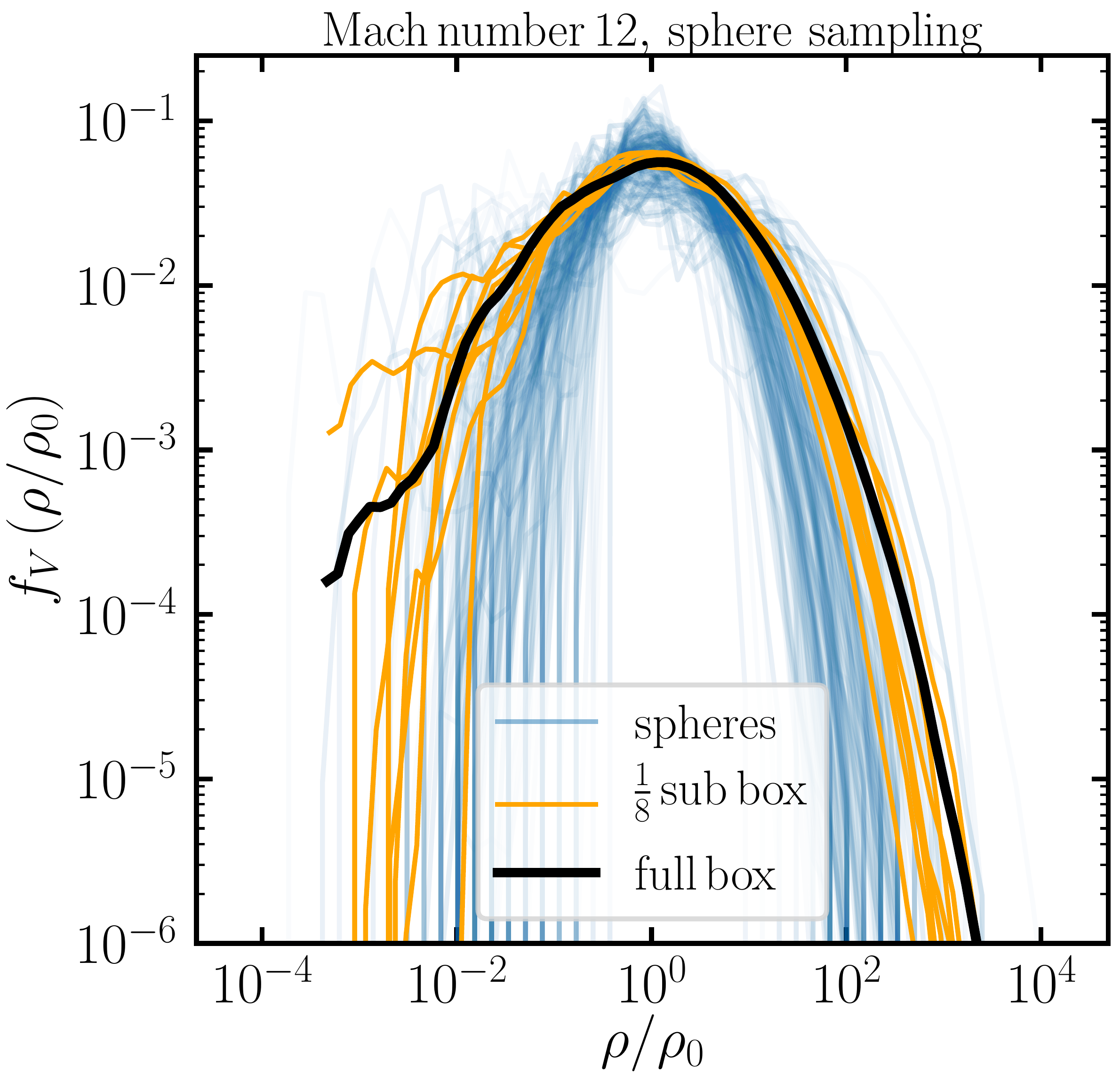}
  \oscaption{density_pdf_M04_0150}{Volume weighted density PDF for turbulent simulation boxes of Mach numbers $\mathcal{M}=4$ and $\mathcal{M}=12$. In the two left-hand panels, the thick black lines show the density PDF of the entire simulation box while orange and blue lines show the density PDF of 8 and 64 disjoint sub-boxes, respectively. The red dashed line shows a lognormal fit to the density PDF of the entire simulation box shown with the black solid line. In the right-hand panel, we compare for the $\mathcal{M}=12$ box the density PDF of individual spheres of radius $R=0.08L_{\rm{box}}$ (blue lines) to the density PDF of the full box (thick black line) and the 8 sub-boxes (orange lines). Each PDF is rescaled to the average density, $\rho_0$, of its respective volume.}
  \label{fig:turbulent_pdf}
\end{figure*}

\begin{figure*}
  \centering
    \includegraphics[width=.33\textwidth]{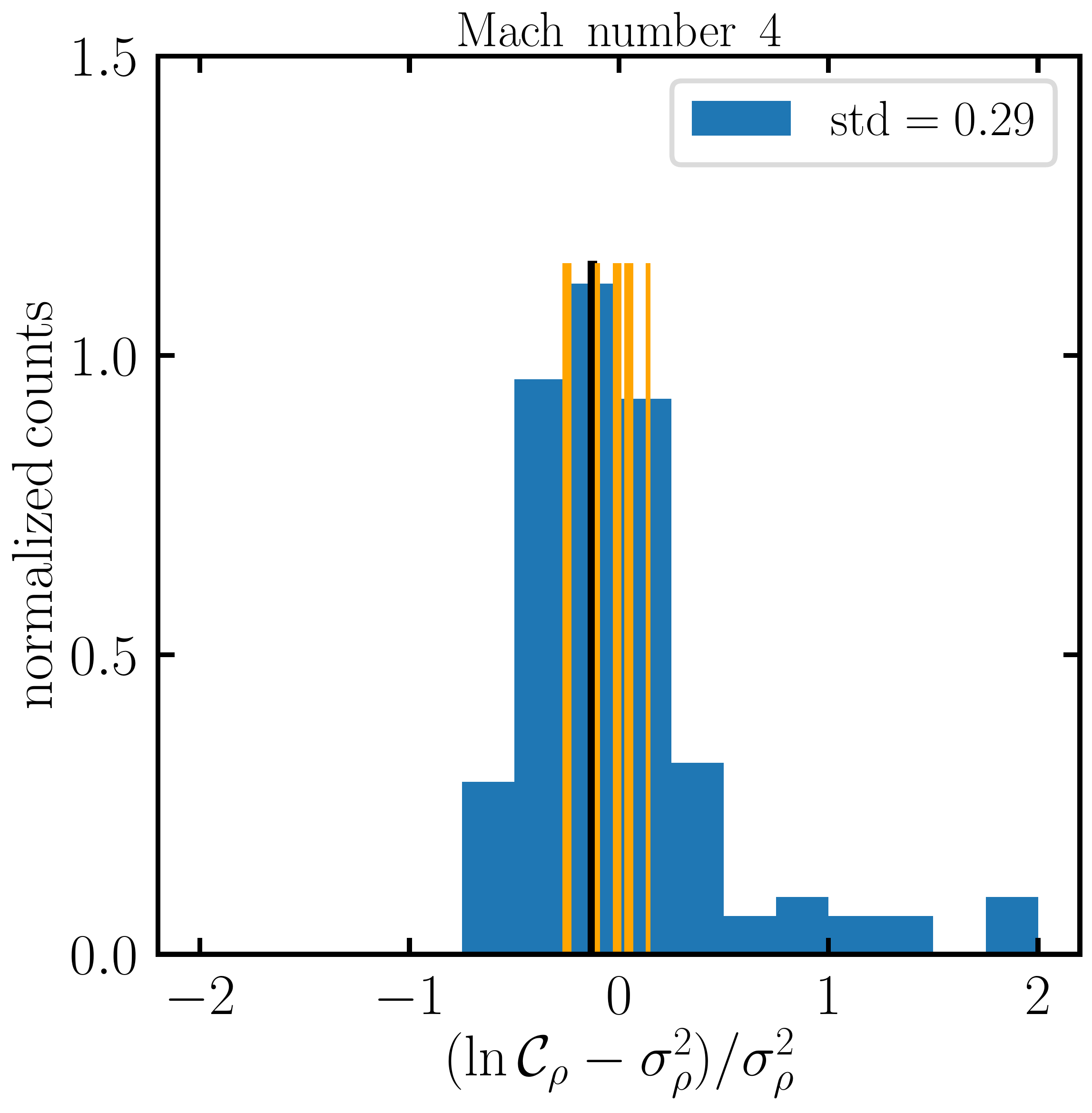}
  \includegraphics[width=.33\textwidth]{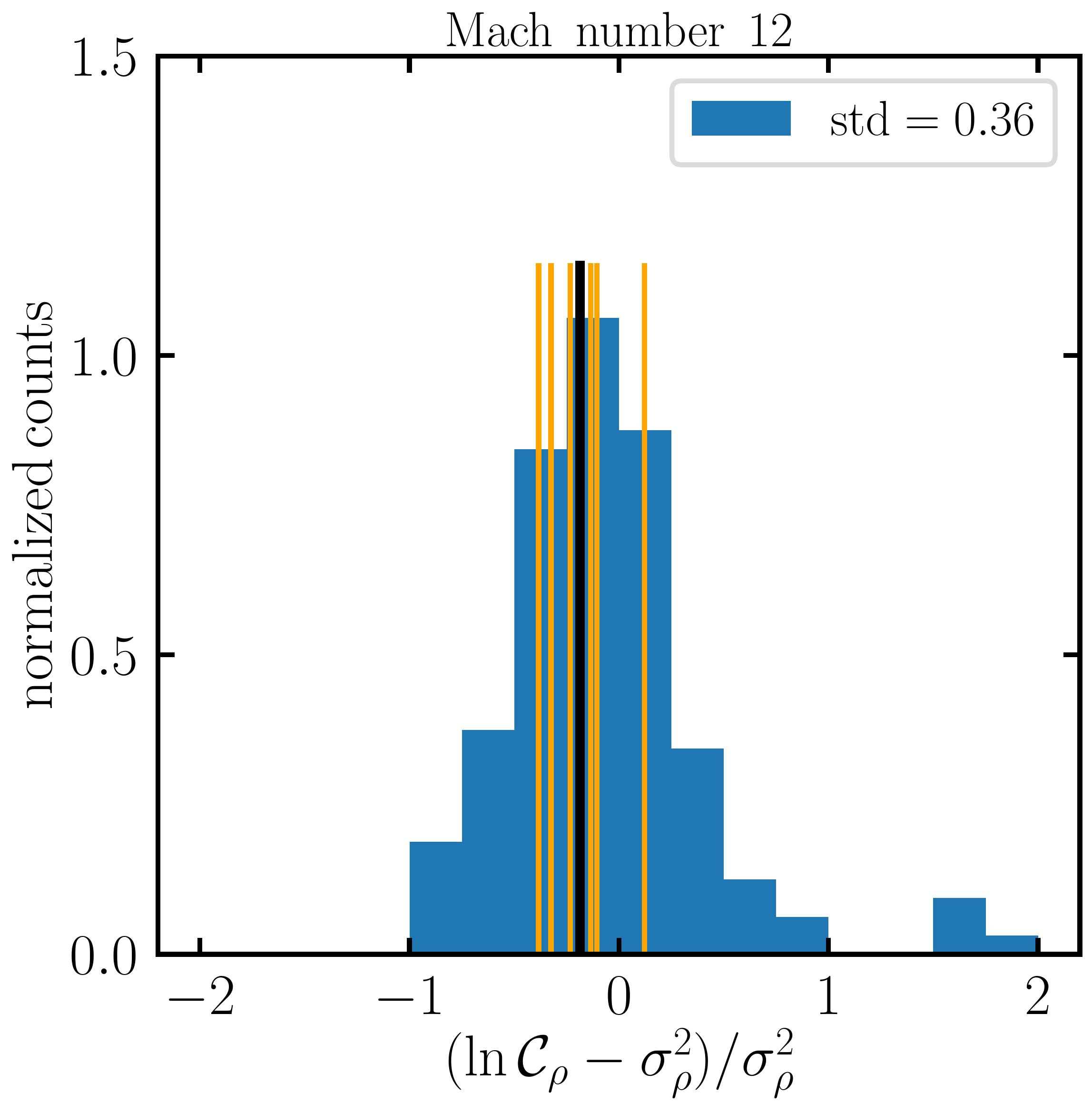}
  \includegraphics[width=.33\textwidth]{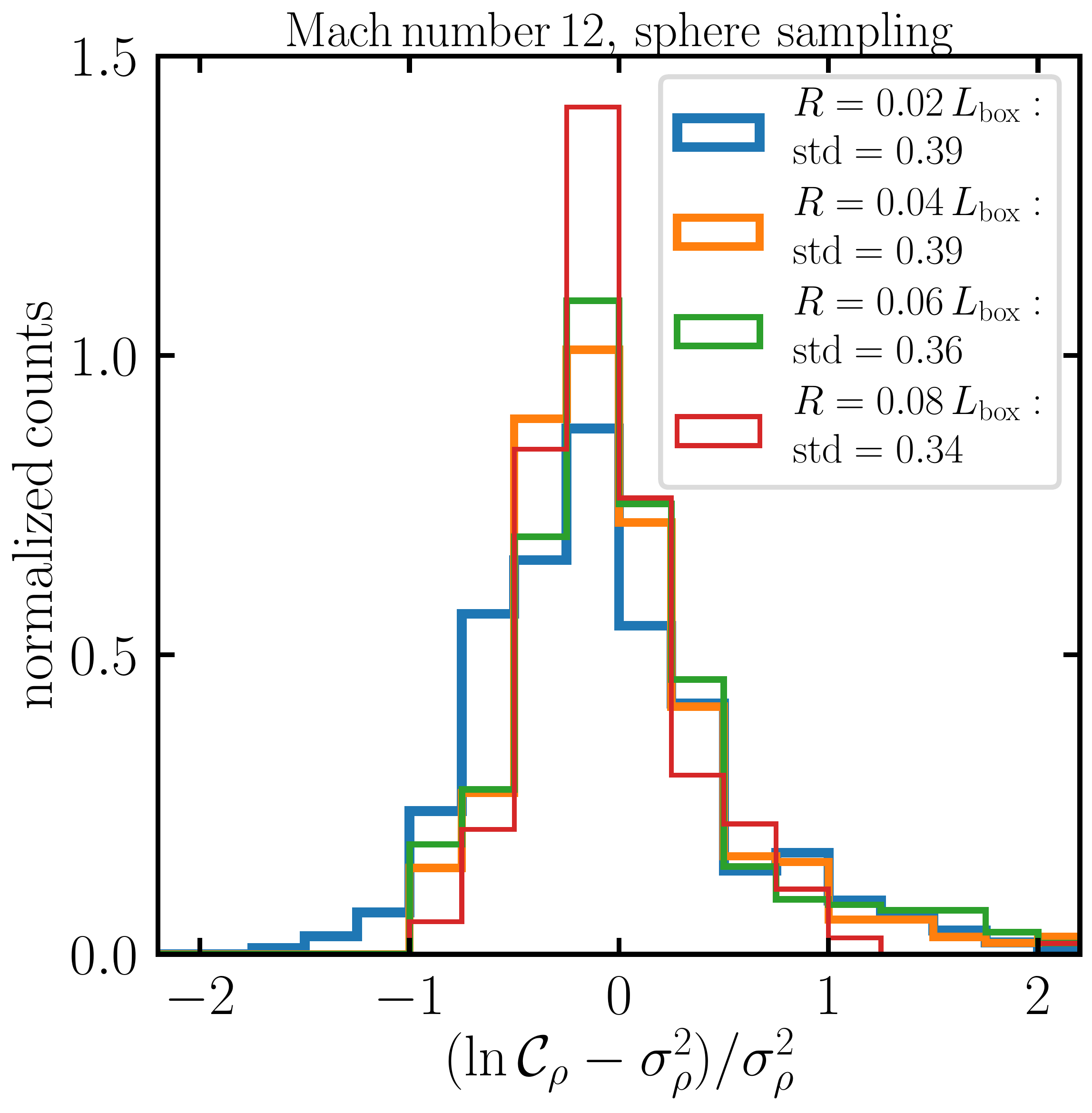}
  \oscaption{error_pdf_M04}{Relative deviation between the logarithm of the clumping factor, $\ln\C_\rho$, and the width of the lognormal density PDF, $\sigma^2_\rho$. The two left-hand panels show the distribution of the relative difference for the simulations with differing Mach numbers. We show the 64 sub-boxes with blue histograms, the 8 sub-boxes with vertical orange lines and the result of the full box with a vertical thick line. The right-hand panel shows the relative difference between the logarithm of the clumping factor and the density PDF width for spheres of different radii for the case of Mach number $\mathcal{M}=12$. According to Eq.~\eqref{eq:clump}, the relative difference should be zero for a perfect lognormal density distribution.}
  \label{fig:error_hist_turb}
\end{figure*}

\begin{figure*}
  \centering
  \includegraphics[width=.8\textwidth,trim={245 0 0 0 cm},clip]{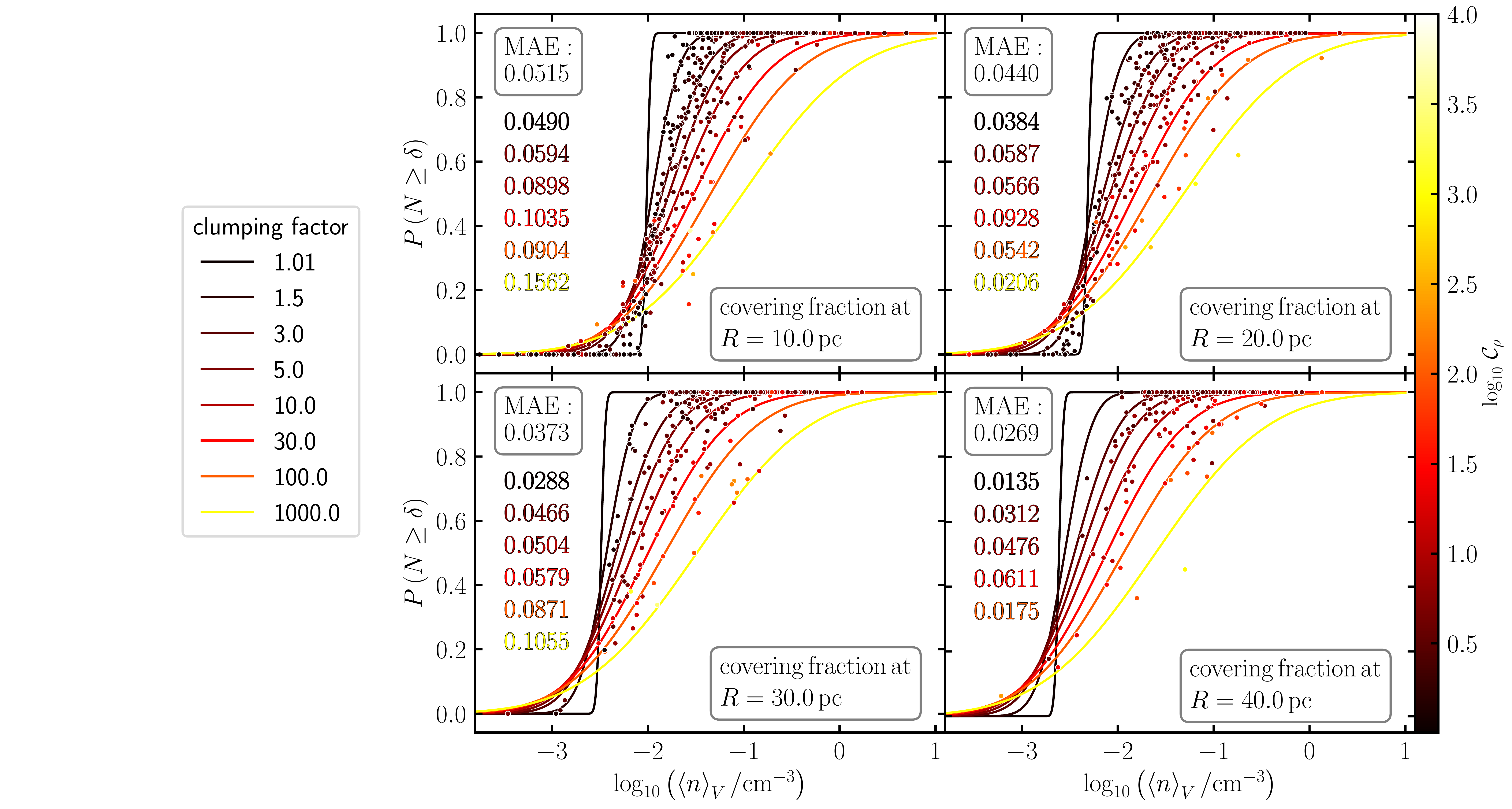}
  \oscaption{theory_vs_turbulent_data_2}{Comparison of the theoretical expectation (lines) for the covering fraction as a function of density with data from turbulent boxes of Mach number $\mathcal{M}=12$ (colored dots) for fixed column density threshold of $\delta=10^{17}$ cm$^{-2}$. Each dot in the four different panels represents one sphere of radius $R$ and we show increasing sphere radii from left to right as indicated in the lower right text box. The color-coding shows the clumping factor and colored numbers on the left of each panel show the mean absolute error (MAE) between the theoretical covering fraction (using Eq.~\ref{eq:err3}) and the one measured from the simulation in six bins of clumping factor (bin edges $\C_\rho = [1,5,10,30,100,1000,10000]$). The upper left text box denotes the MAE of all data points.}
  \label{fig:cov_frac_turb}
\end{figure*}

A lognormal density PDF is only realized in isothermal, supersonic  turbulent gas. However, deviations at the smallest and largest densities can occur as a result of (temporal and/or spatial) correlations and intermittency. Once self-gravity is included, the collapse and fragmentation of the highest density peaks results in a power-law tail of the density PDF \citep[e.g.][]{Girichidis2014}. Thus, in order to validate our model we first look at such idealized, driven turbulence simulations for which all assumptions in Section~\ref{sec:lognorm} are fulfilled. In Figs.~\ref{fig:turbulent_pdf} and \ref{fig:error_hist_turb} we first check how well turbulent patches of gas are described by a lognormal density PDF and how well Eq.~\eqref{eq:clump} determines the width of the distribution functions. Following this, in Fig.~\ref{fig:cov_frac_turb} we investigate how well our final result of Eq.~\eqref{eq:err3} works in this idealized, turbulent regime.

\subsubsection{Density distribution function}

As stated in Section \ref{sec:lognorm} for purely supersonic turbulence the density distribution has a lognormal shape where the width is related to the Mach number (see Eq.~\ref{eq:mach}). Figure~\ref{fig:turbulent_pdf} shows the density distribution for a turbulent box simulation of Mach number $\mathcal{M}=4$ (left panel) and $\mathcal{M}=12$ (middle and right panels). The black line shows the density distribution for the entire box while orange and blue lines show the density PDF of 8 and 64 disjoint sub-boxes when cutting the simulation domain in in half (quarter) in each dimension. In the right-most panel black and orange lines are the same as in the middle panel while the blue lines now show 500 randomly sampled spheres of radius $R=0.08L_{\rm{box}}$.

Figure~\ref{fig:turbulent_pdf} indeed shows that the density PDF for turbulent boxes follows a lognormal shape and the width of the density PDF of Mach number $\mathcal{M}=4$ is narrower compared to the one of Mach number $\mathcal{M}=12$, as expected. From this figure we further see that the density PDF is deviating from a purely lognormal shape (shown in red dashed) at lower and high densities, which is caused by large-scale density correlations in both under- and overdense regions.
Furthermore, we find that restricting to sub-volumes creates deviations from the PDF of the entire box and sampling effects imprint themselves on the density PDFs. Still, Fig.~\ref{fig:turbulent_pdf} shows that in each case the shape of the density PDF is given by a lognormal with comparable width to the full simulation volume. The deviations between the different PDFs are a result of sampling effects at the wings of the distribution, i.e., the lowest and highest densities realized in the volume under consideration. This sampling effect becomes more prominent the smaller the region of interest and thus, at fixed resolution the smaller the number of sampling points, i.e. simulation cells, gets. Thus, the scatter for the 64 sub-boxes is larger than that of the 8 sub-boxes and even more so the scatter for the 500 random spheres of radius $R=0.8L_{\rm{box}}$. In order to quantify the impact of this effect on our assumption, we fit a Gaussian curve in logarithmic density to each PDF and determine its width $\sigma^2_\rho$. Additionally, we use the definition of $\mathcal{C}_\rho$ (as laid down on the left-hand side of Eq.~\ref{eq:clump}) to calculate the clumping factor and compare its logarithm to $\sigma^2_\rho$. In case of a purely lognormal distribution, we expect those two values to be the same but as we have seen. Density correlations and thus deviations from the central limit theorem will lead to slight deviations from a lognormal shape which we quantify below.

In Fig.~\ref{fig:error_hist_turb} we show the resulting relative error between $\ln\C_\rho$ and $\sigma^2_\rho$ for the same turbulent boxes as shown in Fig.~\ref{fig:turbulent_pdf}. For the entire simulation box, the clumping factor, $\C_\rho$, determines the width of the PDF to better than 12 per cent at Mach number $\mathcal{M}=4$ and better than 18 per cent at Mach number $\mathcal{M}=12$ (black vertical lines). Similarly, for the 8 sub-boxes, $\C_\rho$ determines the width of the PDF to better than 25 per cent at $\mathcal{M}=4$ and better than 38 per cent at $\mathcal{M}=12$ (orange vertical lines). For the 64 sub-boxes we show the full error distribution in blue bars which has a standard deviation of 29 per cent (36 per cent) at Mach number $\mathcal{M}=4$ ($\mathcal{M}=12$). The right-most panel of Fig.~\ref{fig:turbulent_pdf} shows the error distribution of 500 random spheres of different radii sampled from the simulation box which also show standard deviations of $\sim30$--$40$ per cent. 

From this analysis we conclude that on average Eq.~\eqref{eq:clump} is indeed fulfilled when sampling patches from idealized turbulent simulations boxes. However, we also find that sampling effects and large-scale correlations may compromise the accuracy of Eq.~\eqref{eq:clump}. For our tests performed here, we expect $\sim30$ per cent uncertainty when using the clumping factor to determine the width of the density PDF. In the next subsection we quantify how this uncertainty on $\sigma^2_\rho$ propagates through Eq.~\eqref{eq:err3} to the estimated covering fraction.

\subsubsection{Covering fraction}
In order to derive the covering fraction following Eq.~\eqref{eq:err3} we scale our turbulent box simulation to a side length of $L_{\rm{box}}=500$ pc and a mean density of $\langle{n}\rangle_V=0.1$ cm$^{-3}$. Then we sample 500 random spheres for 4 different radii from the simulation domain and calculate the gas clumping factor in each sphere as well as its covering fraction when adopting a threshold value of $\delta=10^{17}$ cm$^{-2}$. Figure~\ref{fig:cov_frac_turb} shows the results of this analysis. Each dot represents one sphere with the color code highlighting the clumping factor of that sphere and colored lines show the result of Eq.~\eqref{eq:err3} for fixed clumping factor. Additionally, for each sphere we use its clumping factor as an input to Eq.~\eqref{eq:err3} to calculate its theoretical covering fraction. The mean absolute error (MAE, $\sum_{i=0}^{N} \vert P_{\rm{theo}} - P_{\rm{sim}}\vert / N$) between this theoretical covering fraction and the one measured from the simulation in bins of clumping factor (bin edges $\C_\rho = [1,5,10,30,100,1000,10000]$) is shown with colored numbers in the left part of each panel. In general, we find that our model shows small MAE of the covering fraction of less than 0.1 for most clumping factor bins (except for the smallest sphere radii and the largest clumping factors). In most cases the error is even smaller than 0.05. In general, the model error is lowest for the smallest clumping factors and further decreases with increasing sphere radius. We have further explored how the error depends on average sphere density but found no strong dependence, although we note that around the sharp, step-like increase in covering fraction the uncertainty of the model peaks. Thus, we conclude that although sampling effects might play a role for the exact determination of the density PDF width, their effects on the final covering fraction is small. Therefore, we continue to apply our model to more realistic, full-physics applications in the next sub-section.

\subsection{SILCC simulations of the solar neighbourhood} \label{sec:sim}

\begin{figure*}
  \centering
  \includegraphics[width=\textwidth]{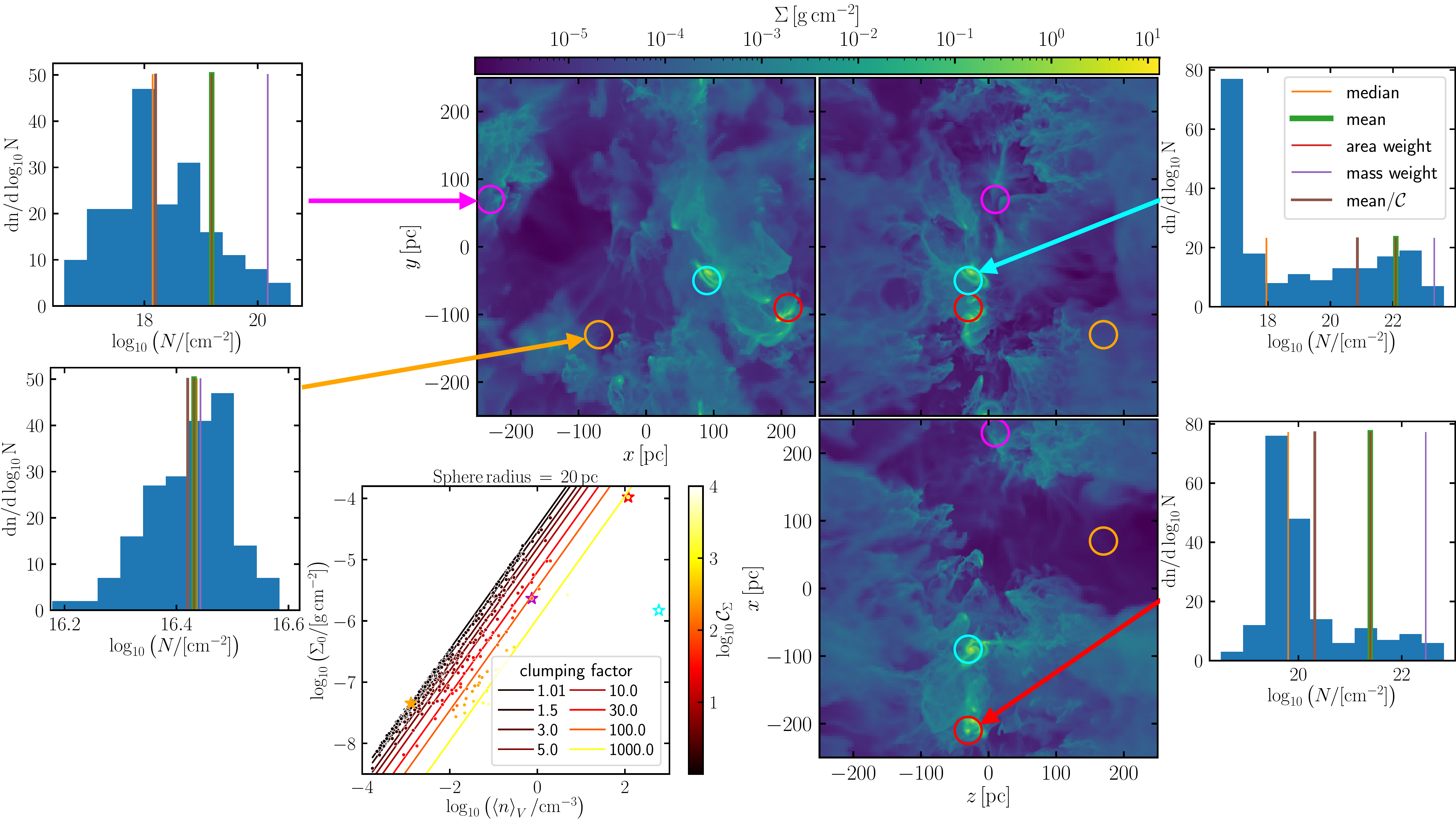}
  \oscaption{comparison_plot_brand_new}{ Illustration of the impact of the clumping factor on the column density distribution in the SILCC simulation. The histograms show the column density distribution of four example spheres of radius $R=20$ pc at different locations in the ISM as shown with colored circles in the surface density plot in the center panels (the three panels show the different spatial projections along the three coordinate axes). The two histograms on the left-hand side show more uniform, low density environments while the two histograms on the right-hand side show highly clumped environments with a large mean density. The lower left panel displays the relation between peak column density $\Sigma_0$ and average three dimensional mass-weighted density, $\langle\rho\rangle_M$, of all spheres with radius $R=20$ pc shown with colored dots (see Section \ref{sec:data} for explanation). Colored lines highlight the theoretical relation from Eq.~\eqref{eq:sigma0}. For both, the lines and the points, the color-coding gives the value of the clumping factor.}
  \label{fig:illustration}
\end{figure*}

The SILCC simulations correspond to a segment of a typical galactic disc at low redshift with solar neighborhood properties. The simulation domain for the higher resolution setups in \citet{GirichidisEtAl2018b} focuses on the dense gas structures and covers is a $(500~\rmn{pc})^3$ with a vertical stratification and the calculations are performed with the 3D magneto-hydrodynamic (MHD), adaptive mesh-refinement (AMR) code \texttt{FLASH} in version 4 \citep{Fryxell2000,Dubey2008}. In order to obtain an accurate picture of the ISM, the simulations include an external galactic potential, self-gravity, radiative heating and cooling, chemical evolution (which follows the formation of $\rm{H}_2$ and CO molecules) with non-equilibrium abundances as in \citet{NelsonLanger1997}, \citet{Glover2007} and \citet{Micic2012}, supernova feedback for both isolated and clustered SNe as well as magnetic fields \cite[for a complete description of the simulation setup and physics implementation see][]{Walch_2015,Girichidis2016}.

We perform our analysis for five different sphere radii ranging from 10 to 50 pc in increments of 10 pc. These scales are representative for gas clouds including their immediate vicinity, in which stellar feedback acts. These radii approximately corresponds to current resolutions of cosmological Milky Way simulations \citep[e.g.][]{Grand2017,Buck2020,Applebaum2021,Agertz2021}. For each radius we sample a total of 500 randomly chosen spheres from the simulations. 
For each sphere we calculate its volume-averaged density $\langle\rho\rangle_V$ by simply summing over all cell masses $m_i$ inside the sphere and dividing by its volume.

A visual representation of typical sphere positions and their corresponding column density distribution is given in Fig. \ref{fig:illustration}. The four example spheres of radius $R=20$~pc highlighted in this figure show the diversity of column density distributions resulting from different environments sampled from the simulation. This figure clearly shows that the average density is the main factor for differences in the column density distribution. The reason for this is that higher density regions are statistically more structured than low density regions which exhibit a more uniform density distribution.

\subsubsection{ISM clumping in different environments}
\label{sec:sampling}

The model we derived in Section \ref{sec:theo} has three free parameters:, $R$, $\C_\rho$ and $\langle\rho\rangle_V$. Because this model is developed to estimate unresolved small-scale ISM structures in low-resolution simulations, the size and density of gas clouds are immediately identified with the size and density of the resolution elements in the low-resolution simulation.

Thus, the only real free parameter for the theoretical model is the clumping factor, $\C_\rho$, of the ISM. Choosing a value for $\C_\rho$ the sub-grid structure of a given patch of the ISM (the gas density distribution and the column density distribution and hence the covering fraction) is solely determined by the average ISM density of that region. Additionally, the average density of a gas parcel is exactly what coarse resolution simulations trace for all their resolution elements. Therefore, in order to establish a statistical model for the ISM which intersects self-consistently with coarse resolution simulations we need the (statistical) connection between the clumping factor and the average density. 

\begin{figure}
  \centering
  \includegraphics[width=\columnwidth]{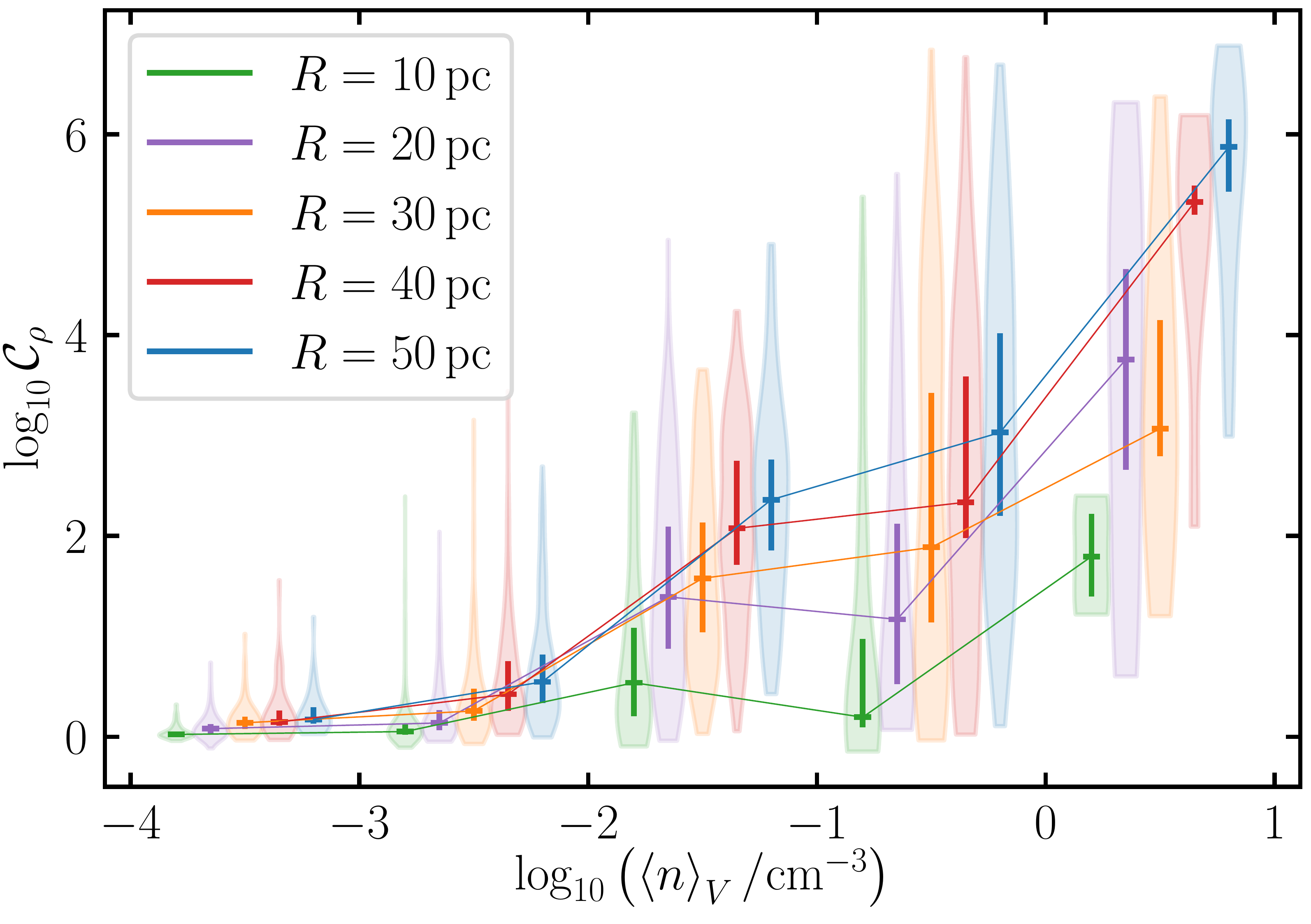}
  \oscaption{c_vs_rho}{ISM density clumping factor as a function of density for five different sphere radii as derived from the SILCC simulation. We show the full distribution of clumping factors in $1$~dex wide bins of density calculated via a kernel density estimate. The individual violins extend to the minimum/maximum value in each bin. The vertical error-bars inside each violin show the scatter around the median calculated as the $32^{\rm{nd}}$ and $68^{\rm{th}}$ percentile. For better visibility we have slightly offset each violin for different averaging radii.}
  \label{fig:c_vs_rho}
\end{figure}

Because of the above mentioned difficulties and to be flexible to adapt to new theoretical insights, we have established a phenomenological connection between the clumping factor and the ISM density from the SILCC simulations in Fig.~\ref{fig:c_vs_rho}.
This figure uses the clumping factor defined in Eq.~\eqref{eq:clump} to describe the structure of the ISM and
shows the median clumping factor as well as its scatter (calculated as the $32^{\rm{nd}}$ and $68^{\rm{th}}$ percentile) as a function of the average ISM density $\langle\rho\rangle_V$. With different line colors we show results for the different sphere radii tested here. This figure echos our qualitative findings from the previous section. 
Figure~\ref{fig:c_vs_rho} clearly shows that for each radius probed in this work, the clumping factor increases with increasing ISM density, $\langle\rho\rangle_V$. For ISM densities below $\langle\rho\rangle_V\lesssim10^{-2}\,\mathrm{cm}^{-3}$ the clumping factor is essentially equal to unity while for densities larger than this it rises quickly towards values around 10-100.  
Thus, the higher the ISM density, the more substructure we expect to find.
Numerical values for the median and scatter in each density bin are shown in Tab. \ref{tab:clump}.

\begin{table}
\begin{center}
\caption{Clumping factor distributions as a function of density and sphere radius as derived from the SILCC simulations. For each radius we state the median, 32$^{\rm nd}$ and the 68$^{\rm th}$ percentile of the clumping factor for a range of $1$~dex wide ISM density bins.}
\label{tab:clump}
\begin{tabular}{ l c c c c c}
  \hline  \hline
  \addlinespace[1ex]
  \multicolumn{2}{l}{$R$ [pc]}& \multicolumn{3}{c}{$\log_{10}\left(\frac{\rho}{\rm{cm^{-3}}}\right)$} & \\ 
  \addlinespace[1ex]
  \hline
  \addlinespace[.5ex]
	& $(-4,-3)$ & $(-3,-2)$ & $(-2,-1)$ & $(-1,0)$ & $(0,1)$ \\
	\addlinespace[.5ex]
    \hline
    \addlinespace[.5ex]
    10	& 0.02$_{0.01}^{0.05}$ & 0.05$_{0.02}^{0.12}$ & 0.54$_{0.2}^{1.08}$ & 0.19$_{0.09}^{0.97}$ & 1.79$_{1.39}^{2.22}$ \\
  	\addlinespace[1ex]
	20  & 0.08$_{0.03}^{0.13}$ & 0.14$_{0.06}^{0.27}$ & 1.39$_{0.87}^{2.09}$ & 1.17$_{0.52}^{2.12}$ & 3.76$_{2.66}^{4.66}$ \\
	\addlinespace[1ex]
	30	& 0.14$_{0.08}^{0.2}$ & 0.26$_{0.16}^{0.48}$ & 1.58$_{1.04}^{2.14}$ & 1.89$_{1.14}^{3.42}$ & 3.07$_{2.79}^{4.15}$ \\
	\addlinespace[1ex]
	40	& 0.15$_{0.11}^{0.26}$ & 0.42$_{0.26}^{0.75}$ & 2.08$_{1.71}^{2.75}$ & 2.34$_{1.98}^{3.59}$ & 5.32$_{5.2}^{5.49}$ \\
	\addlinespace[1ex]
	50  & 0.17$_{0.13}^{0.29}$ & 0.55$_{0.33}^{0.82}$ & 2.36$_{1.86}^{2.76}$ & 3.03$_{2.2}^{4.02}$ & 5.87$_{5.43}^{6.15}$ \\
	\addlinespace[.5ex]
  \hline
\end{tabular}
\end{center}
\end{table}

\begin{figure*}
  \centering
  \includegraphics[width=\textwidth]{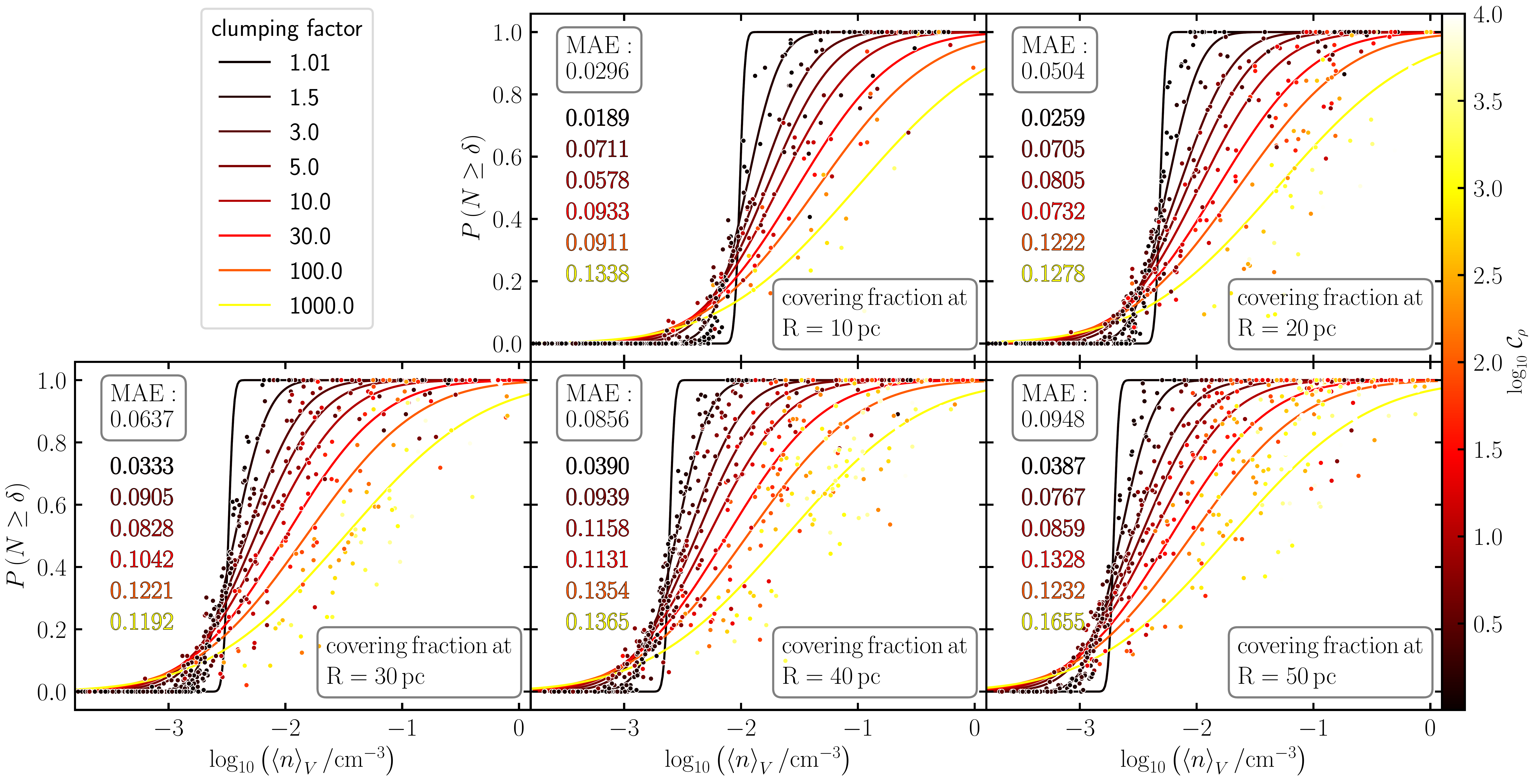}
  \oscaption{theory_vs_data2}{Comparison of the theoretical expectation (lines) for the covering fraction as a function of density with data from the SILCC simulations (colored dots)  for fixed column density threshold of $\delta=10^{17}$ cm$^{-2}$. The five different panels show increasing sphere radii from left to right as indicated in the text box. Again, the color-coding shows the clumping factor and colored numbers on the left of each panel show the mean absolute error between measured covering fraction and the one derived using Eq.~\eqref{eq:err3} and the measured clumping factor $\C_\rho$ of each sphere in six bins of clumping factor with bin edges $\C_\rho = [1.,5.,10.,30.,100.,1000.,10000.]$. The upper left text box denotes the MAE of all data points.}
  \label{fig:data}
\end{figure*}

Furthermore, we see that there is a secondary correlation of the clumping factor with sphere radius. Larger sphere radii exhibit larger clumping factors independent of ISM density. This is simply due to the fact that the larger the region across which the density contrast is measured, the larger the possible fluctuations become. This property assumes a decreasing ISM density power spectrum with scale and is realized for Kolmogorov or Burgers turbulence as long as the injection scale is larger than the averaging scale.

Figure~\ref{fig:c_vs_rho} enables us to statistically sample valid clumping factors as input to our model \citep[see also Section 2.3 III of][for a similar approach to sub-grid IGM clumping]{Bianco2021}. With this approach we are then able to derive the sub-grid structure of the ISM gas and corresponding covering fractions as outlined in Section~\ref{sec:method}.

\subsubsection{Covering fractions of different environments}

Finally, we gauge the validity of our model by comparing theoretical predictions for covering fractions to the results from the SILCC simulations. Figure~\ref{fig:data} shows the covering fractions of patches of the ISM as a function of their average density, $\langle\rho\rangle_V$. From left to right, this figure shows the results for five different sphere radii as indicated by the panel's text box. Coloured lines show the results from our theoretical model for different clumping factors, $\C_\rho$ and coloured points show the covering fractions calculated for spherical regions of the ISM in the SILCC simulations where the colour-coding shows the clumping factor, $\C_\rho$, of the simulated ISM patches. For the 5 values of sphere radii probed in this work the correspondence between the model and the data is reasonably good and data points of a given clumping factor, $\C_\rho$ follow the theoretically predicted curves of the same clumping factor. Our model works best for radii between $R=20$ to $40$ pc while for the smallest and the largest radii we note some deviations of the simulated points from the theoretical predictions. Similarly, the MAE in bins of clumping factor shows that the model predictions are best for smaller clumping factors below a value of $30$--$100$. Above those values of $\C_\rho$ physical effects cause deviations from the lognormal assumption and lead to model deviations. In general, the outlier points highlight how the simplified assumptions of a log-normal PDF (see also discussion of Fig.~\ref{fig:illustration}) and adopting the mean relation, $\C_\rho=\C_\Sigma^{2/3}$ (Eq.~\ref{eq:C_rho_vs_C_Sigma}), affects the results. As stated above, a possible solution to this simplification is to extend the model to include more elaborate gas density PDFs, e.g., a log-normal PDF with a power-law tail towards large densities and/or to additionally include the scatter in the relation between $\C_\rho$ and $\C_\Sigma$ from Fig.~\ref{fig:c_vs_c}.

\subsubsection{Connecting covering fractions with photon escape fractions}

Photons in the ISM escape through low density channels and there exist a clear correlation between e.g. Lyman continuum radiation and HI column density \citep[e.g.][]{Kakiichi2019}.
Figure 8 of \citet{Kakiichi2019} shows how the escape fraction of Lyman continuum radiation depends on the covering fraction in radiation hydrodynamical simulations. Their analysis reveals a near linear dependence of escape fractions on covering fractions with slopes of $\sim0.4$ and $\sim0.6$ depending on velocity dispersion. Combining our model for the covering fraction of ISM gas clouds with their results for the dependence of escape fractions on covering fractions it is straight forward to derive Lyman continuum or Lyman-$\alpha$ escape fractions for gas clouds in the ISM.

\section{Discussion}
\label{sec:appl}

Most galaxy-scale and all cosmological simulations of galaxies with halo masses $M\gtrsim10^{11}\,\rmn{M}_\odot$ lack the resolving power to model the detailed physics of the star formation and feedback cycle. Not only is sufficient spatial resolution crucial for an accurate and correct treatment of the necessary physics, but also the complexity of the physical processes of star formation \citep[see e.g.][for reviews on star formation processes]{Girichidis2020,Ward-Thompson2015} and feedback \citep[e.g.][]{Krause2020} prevent an ab-initio treatment of these effects in larger galaxies. Therefore, galaxy formation simulations commonly take an agnostic approach to star formation and feedback by implementing parametric models \citep[see e.g.][for recent reviews]{Somerville2015,Naab2017}. The models for star formation typically require a normalization, the star formation efficiency, $\epsilon_{\star}$, as well as a parameter determining the scaling with gas density. Similarly, coupling the (stellar) feedback energy to the ISM requires a coupling efficiency, $\epsilon_{\rm{FB}}$. Both parameters are adjusted to match the zero point and the slope of the observed Kennicutt-Schmidt relation \citep{Schmidt1959,Kennicutt1998} between star formation rate surface density and gas surface density as well as other observed galaxy scaling relations.

Here, we presented a new framework to model unresolved or uncertain physics with a statistical approach. We have built a statistical sub-resolution model which at a given spatial scale $R$ solely depends on the value of the gas clumping factor, $\C_\rho$, at the average density of that resolution element. The implicit dependence of $\C_\rho$ on density encodes the unknown or uncertain physical mechanisms that shape the density structure below the resolution limit of the low-resolution simulation. The connection between $\C_\rho$ and average density, $\left<\rho\right>_V$ on a given scale $R$ can be derived using high-resolution, high-fidelity simulations of the ISM.

In this section we highlight some areas of possible applications of the model to improve the current sub-grid models in coarse-resolution simulations of galaxy formation. We caution that adequately addressing each of those applications is much more complex than outlined here and certainly deserves its own dedicated research program to be accurately modelled in galaxy scale simulations. Thus, we leave an extension of the model concepts presented here to more complex setups for future work.

In the following three sub-sections, we exemplify how a statistical treatment of ISM sub-grid clumping can be applied to model star formation, (stellar) feedback and attenuation of stellar radiation. Finally, with slight modifications the framework presented here can also be used to measure the star formation efficiency, $\epsilon_\rmn{ff}$, \citep[similar to][]{Hu2021b} by improving the estimates of the cloud's free-fall time from its volume-average mean density, $\left<\rho\right>_V$, including the effects of cloud sub-structure \citep[see also][for a similar formalism]{Hu2021}.    

\subsection{Cold, dense gas and star formation}

The basic recipe for star formation in many simulations of galaxy formation still follows the pioneering work of \citet{Katz1992}. Dense and converging gas is assigned a SFR based on a \citet{Schmidt1959} law:
\begin{align}
    \dot\rho_\star=\frac{\epsilon_\star\rho_{\rm{gas}}}{t_{\rm{ff}}}\propto\rho_{\rm{gas}}^{1.5},
\end{align}
where the proportionality in the last step follows from the proportionality of the local free-fall time with gas density $t_{\rm{ff}}\propto\rho_\rmn{gas}^{-0.5}$. The star formation efficiency parameter, $\epsilon_{\star}$, is typically calibrated to match the amplitude of the observed \citet{Kennicutt1998} relation.

In the star formation community, however, different star formation criteria have been put forward for calculating the fraction of star-forming mass from considerations of the turbulent structure of molecular clouds \citep[e.g.][]{Padoan1995,Hennebelle2008}, the core mass function of fragmenting gas clouds \citep[e.g.][]{Padoan2002,Banerjee2014,Voelschow2017} or simply the mass fraction of dense gas \citep[e.g.][]{Elmegreen2002,Krumholz2005} irrespective of the mass function of dense cores. The reasoning in this work closely follows those latter considerations.

Since we are not interested in modelling the mass function of newly forming stars, for our purpose,  $\epsilon_{\star}$ measures the fraction of (dense gas) mass of the simulation's resolution element that is eligible to be converted into stars in a given timestep. This approach most closely follows the logic of \citet{Krumholz2005}. Typically, models take the amount of gas above a given threshold density $\rho_{\rm crit}/\rho_0$ to represent the fraction of gas in self-gravitating and collapsing gas clouds. More sophisticated models use the virial parameter, the ratio between kinetic and potential energy of a gas cloud, to localize star forming regions \citep[e.g.][]{Semenov2019} or have incorporated sub-grid recipes to compute the density of molecular hydrogen $\rho_{\rm H_2}$ to replace the arbitrary density threshold \citep[e.g.][]{Kuhlen2012,Christensen2012,Agertz2015,Lupi2018}. However, it is still unclear whether $\rmn{H}_2$ formation is the primary driver for star formation \citep[see e.g.][]{Glover2012,Krumholz2013}.

While our model is similar to models of $\rmn{H}_2$-based star formation, it circumvents the need for numerically resolving the physics of the complex $\rmn{H}_2$ formation processes and estimates the star formation efficiency, $\epsilon_{\star}$, simply as the mass fraction of dense gas of a coarse resolution element.
Equation \eqref{eq:clump} in combination with Eq.~\eqref{eq:cumulative} allows us to equate the fraction of dense gas as a function of $\rho_{\rm crit}$ and clumping factor $\C_\rho$:  
\begin{align}
P\left(\rho\geq\rho_{\rm crit}\right) &= \frac{1}{2}\left(1-\mathrm{erf\left[\frac{\ln\left(\rho_{\rm crit}/\rho_0\right)}{\sqrt{2 \ln\C_\rho}}\right]}\right) \equiv \epsilon_{\star}
\end{align}
For suitable choices of $\rho_{\rm crit}/\rho_0$ and appropriate clumping factors $\C_\rho$, either via Eq.~\eqref{eq:mach} \citep[see e.g.][for a recent observational determination of the turbulent driving parameter]{Sharda2021} or via statistically sampling it following Fig.~\ref{fig:c_vs_rho} this equation immediately yields the star formation efficiency $\epsilon_\star$ \citep[see also Section 2.2 of][for a similar approach to model the star formation efficiency]{Lupi2018}.

\subsection{Boosting of stellar winds and supernovae explosions}

The expansion of spherical shock fronts such as supernova explosions or stellar wind bubbles strongly depends on the ambient medium \citep[e.g.][]{Kim2015,Steinwandel2020,Lancaster2021a}. Furthermore, the generation of the expected radial momentum of the resolved shock front depends on the resolution of the simulation \citep[e.g.][]{Gutcke2021}.

The momentum during the Sedov-Tailor (ST) phase scales as $P\propto t^{3/5} \rho^{1/5}$ but in higher density regions the explosion transitions earlier to the momentum conserving phase compared to lower density regions \citep[e.g. figure~3 of][]{Haid2016}. Thus, at fixed time SNe exploding in high density regions develop a larger radial momentum due to the larger swept-up mass in the shell compared to low-density regions. On the other hand, the ST phase lasts longer in low-density regions such that at transition to the momentum conserving phase, the radial momentum is larger in those regions. However, in turbulent molecular clouds, regions of low- and high-density co-exist and thus the momentum generated by SNe exploding in such media will be different from exploding SNe in a uniform media. Especially, the coupling efficiency of the explosion energy with the surrounding gas will depend on the turbulent structure of the ISM via its 3D-density structure.

In general, galaxy simulations lack the resolution to properly resolve the hydrodynamics of exploding stars and a diverse set of sub-grid models have been proposed to circumvent this issue \citep[][]{Somerville2015,Naab2017,Vogelsberger2020}.

One way to improve upon this is by studying supernova explosions in turbulent gas clouds in a highly resolved idealized setup \citep[e.g.][]{Martizzi2015,Pais2018}. For example, \citet{Haid2016} have studied with such idealized simulations how a turbulent medium is able to boost the momentum of exploding supernovae compared to a uniform medium of the same average density. In their Fig.~12 and correspondingly in equation (34) they quantify how the momentum of a single supernova exploding in a turbulent medium of given Mach number is boosted in comparison to an explosion in a uniform medium of the same average density \citep[see also Fig.~17 of][for the retained stellar cluster wind energy in turbulent gas clouds]{Lancaster2021}. Thus, another application of our model is the calculation of the \emph{local, effective} momentum input arising from SN explosions or stellar winds in a non-uniform, turbulent medium. 

The logic here is as follows: For a broader PDF, the density variations become larger and as such, the expanding blast wave encounters more low density regions. Those are subject to less radiative cooling and allow for a higher momentum injection. Next to the turbulent structure of the clouds, the exact quantification of this momentum boost depends on various physical processes such as the propagation of shocks in a clumped medium with radiative cooling \citep{McCourt2018,Gronke2018,Sparre2019,Sparre2020,Li2020}; in particular in the presence of various metal ion species and molecules \citep{Girichidis2021}, magnetic fields and cosmic rays \citep{Pfrommer2017,Pais2018}. Thus, we advocate for the approach of \citet{Haid2016} and \citet{Martizzi2015} who extract the energy and momentum generation of single SN explosions as a function of local gas properties from high resolution simulations that include all necessary physics. While \citet{Martizzi2015} cast their findings only in terms of gas density and metallicity, marginalizing over the turbulent structure of the clouds, \citet{Haid2016} explicitly include a dependence on the turbulent Mach number. Coupling this with our results of Fig.~\ref{fig:c_vs_rho} for the turbulent structure of gas as a function of density then allows for an improved injection scheme for feedback. Note, by sampling the clumping factor for each resolution element at its specific density we are able to take into account the local the turbulent structure of the gas and allow for more stochasticity.

A conceptually different but much more simplified way of employing our model for the injection of momentum and energy into the ISM is given by the calculation of the surface mass density distribution around a source of feedback. High resolution simulations have shown that in a highly clumped medium, the shock propagates around the dense phase in the hot, dilute phase while it stalls in the dense phase because of momentum conservation \citep[see e.g., figure~A1 of][]{Pais2020}. Thus, similar to the calculation of the covering fraction in Eq.~\eqref{eq:err3}, we might define a threshold surface mass density, $\delta_{\rm FB}$, above which the injected supernova or stellar wind energy will cause only little effect. All sight lines with a surface mass density below $\delta_{\rm FB}$ will also have a lower average density along the line of sight and thus allow the shock to travel and break out of the cloud. Sight lines of larger column density will similarly have larger gas densities along the line of sight and thus preferentially absorb the explosion energy and radiate away its energy. Thus, our model enables us to estimate the fraction of surface area of low-density channels through which the shock will escape the cloud. This translates into a coupling efficiency for feedback energy. Working out a suitable value for $\delta_{\rm FB}$ requires high resolution simulations of exploding SNe and stellar winds in turbulent boxes, which is beyond the scope of this work but will be addressed in future work.

\subsection{Attenuation of stellar radiation}

Another source of feedback originates from the radiation of stars. Especially young massive stars are the sources of  an intense radiation field that photo-heats the surrounding high-density gas to a temperature of about $10^4$ K \citep{Stroemgren1939}, drives small-scale winds that reduce the density surrounding exploding stars and thus increases the efficiency of SN feedback \citep[e.g.][]{Stinson2013,Rosdahl2015,Geen2015,Kannan2020b}. At the same time, radiation escaping the star-forming clouds ionizes the surrounding gas and metals in the ISM as well as the circum-galctic medium with the effect of lowering the cooling rates, which in turn reducing the star formation rate in galaxies \citep[e.g.][]{Cantalupo2010,Kannan2014b,Kannan2016,Obreja2019}. Similarly, radiation pressure, both by trapped IR and UV radiation can impart momentum into the ISM, which may help launching large-scale galactic winds \citep[e.g.][]{Murray2011,Emerick2018}. 

Similarly to the \emph{effective} momentum injection by supernovae explosions, also the injection of radiation is affected by the internal density structure of the clouds. Thus the effective \emph{escape} fraction of radiation from the clouds will depend on the relative fraction of low-density channels through which  photons can escape \citep[e.g.][]{Kakiichi2019}. The model presented in this work is ideally suited to calculate the \emph{local, effective} escape fraction from gas clouds as given by eq.~\eqref{eq:err3} and thus provide the means to more accurately couple the radiation from cloud embedded stars to the coarsely resolved ISM in galaxy simulations.

\section{Conclusion} \label{sec:dis}

We set out to theoretically derive a model for the density structure of the interstellar medium with special emphasis on the applicability of the model as a sub-grid prescription of density structures in coarse-grained cosmological simulations. Starting from the simple assumption that most gas in the ISM follows a log-normal density distribution, we derive how the column density distribution of spherically symmetric clouds depends on the average gas density of the cloud. We explicitly incorporate the small-scale gas clumping into our model using the standard definition of gas clumping factor, $\C_\rho$, as given by Eq.~\eqref{eq:clump} which directly relates to the width of the log-normal PDF, $\sigma_\rho$.

Our final result for the covering fraction as a function of ISM density is given by Eq.~\eqref{eq:err3}. In our model, the covering fraction follows an error function (the cumulative function of the log-normal) whose centroid, $\mu_\rho$, and width, $\sigma_\rho$, are modified by the amount of gas clumping characterized by $\C_\rho$. The model presented in this work is derived to estimate small-scale ISM structures in coarse-resolution simulations. Thus, the parameters for the size, $R$ and volumetric density $\langle\rho\rangle_V$ of gas clouds are immediately identified with the size and density of the resolution elements in the coarse resolution simulation. Thus, the only free input parameter to our model is the ISM clumping factor which depends on the physics on cloud scales. Using small-scale ISM simulations this connection can statistically be established. Here, we have used SILCC simulations to derive how $\C_\rho$ behaves as a function of density (see Fig.~\ref{fig:c_vs_rho}).    

While the assumption of a log-normal density PDF for the ISM gives reasonable results when compared to the GMC scales in SILCC simulations (see Fig.~\ref{fig:data}) this assumption might in fact be too simplistic as previous results have shown \citep[e.g.][]{Alves2017,Khullar2021}. With the framework presented here, it is straight forward to replace the log-normal assumption and re-derive the equations for more complicated density PDFs. Similarly, the connection between gas clumping and density will depend on the exact physics modelled, e.g. it is well established that in idealized simulations there is a strong dependence of gas clumping on the Mach number (see Eq.~\ref{eq:mach}).

However, when more physics such as non-equilibrium cooling and feedback by radiation and cosmic rays are considered, the dependence of gas clumping might become more complicated. With the approach chosen here of empirically deriving the gas clumping from small-scale simulations the modelling procedure can easily adapt to new insights from more advanced simulations without changing the model. At the same time, this approach makes it easy to implement the model into coarse-grained simulations and to sample realistic gas clumping factors at runtime.

We summarize the main ingredients of our model as follows:

\begin{itemize}

  \item Under the assumption that the ISM density PDF and the $4\pi$ column density PDF of a gas cloud are well described by a log-normal distribution, it follows that the characteristic densities, $\rho_0$ and widths, $\sigma_\rho$, are functions of the gas clumping factor. We find that the projected column density clumping factor,  $\C_\Sigma$, is the square-root of the three dimensional gas density clumping factor, $\C_\rho$, as shown in Fig. \ref{fig:c_vs_c}.
    
  \item There is a linear relation between the average density and the median of the $4\pi$ column density distribution as seen from the center of that gas cloud in the ISM. In particular, the median column density of clouds and the width of the distribution depends on the clumping factor, $\C_\rho$, as shown in Fig.~\ref{fig:theory}.
  
  \item Defining the covering fraction of a gas cloud as the ratio of number of sight lines above a given density threshold to the total number of sight lines, we derive a functional relation between covering fraction and cloud density in Eq.~\eqref{eq:err3}. Our model follows an error function which reflects our assumption of a log-normal density PDF. The only free parameter of this model is the gas clumping factor at a given cloud density. 
  
  \item We have thoroughly tested our model in the regime where all assumptions are fulfilled, i.e. in purely isothermal, supersonically driven-turbulence simulations (Section~\ref{sec:turb} ) as well as multi-physics simulations from the SILCC project (Section~\ref{sec:sim}). In Fig.~\ref{fig:turbulent_pdf} we find that Eq.~\eqref{eq:clump} has a scatter of $\sim30$ per cent, even in the case of pure turbulence. 
  However, for the turbulent simulations and SILCC, the final mean absolute error between our model as stated in Eq.~\eqref{eq:err3} and simulations is low ($\lesssim0.05$ for turbulence and $\lesssim 0.09$ for SILCC).
  
  \item We have characterized the relation between gas clumping and average ISM density using a set of simulations from the SILCC simulations (Fig.~\ref{fig:c_vs_rho}). We find a strong correlation between average density and clumping factor. This empirically derived relation enables sampling valid values of $\C_\rho$ to model sub-grid density structures in coarse-grained simulations such as cosmological models for the Milky Way.  
    
  \item Gas clumping has a strong effect on the covering fraction at fixed ISM density (see Fig.~\ref{fig:data}). Our model predicts that at a given density the cloud covering fractions can vary between $\sim0.1$ up to $1$. This implies that for a given spatial scale and average  ISM density a gas cloud might be completely opaque to radiation emitted from its center or contrary let all the radiation freely escape, solely dependent on the amount of gas clumping inside the cloud.

  \item Combining our prescription with results from radiative transfer simulations to connect the covering fraction with the escape fraction of photons from gas clouds the model can readily be used to estimate photon escape fractions from embedded sources in the ISM.
  
  \end{itemize}

\section*{Data Availability}
SILLC simulations are publicly available at \url{http://silcc.mpa-garching.mpg.de}. A Jupyter notebook containing all plotting routines and data files can be found here: \codelink \url{https://github.com/TobiBu/ISM_subgrid_clumping.git}

\section*{Acknowledgments}
The authors like to thank Aura Obreja, Keri Dixon and Sven Buder for valuable comments to an earlier version of this draft which helped to improve clarity and readability of the manuscript. We thank the anonymous referee for valuable comments that have improved the quality of this manuscript. TB, CP, and PG acknowledge funding from the European Research Council under ERC-CoG grant CRAGSMAN-646955. PG also acknowledges funding from the ERC Synergy Grant ECOGAL (grant 855130). This research made use of the {\sc{matplotlib}} \citep{matplotlib}, {\sc{SciPy}} \citep{scipy} and {\sc{NumPy, IPython and Jupyter}} \citep{numpy,ipython,jupyter} and {\sc{YT}} \citep{YT} {\sc{python}} packages. Results in this paper have been derived using the \texttt{healpy} \citep{healpy} and \texttt{HEALPix} \citep{Gorski_2005} packages.
Hyperlink figures to code access are inspired by Sven Buder and Rodrigo Luger.



\bibliography{astro-ph.bib}

\appendix

\section{$\C_\Sigma$ vs. $\C_\rho$ relation}
\label{app:c_vs_c}

\begin{figure*}
  \centering
  \includegraphics[width=1.0\columnwidth]{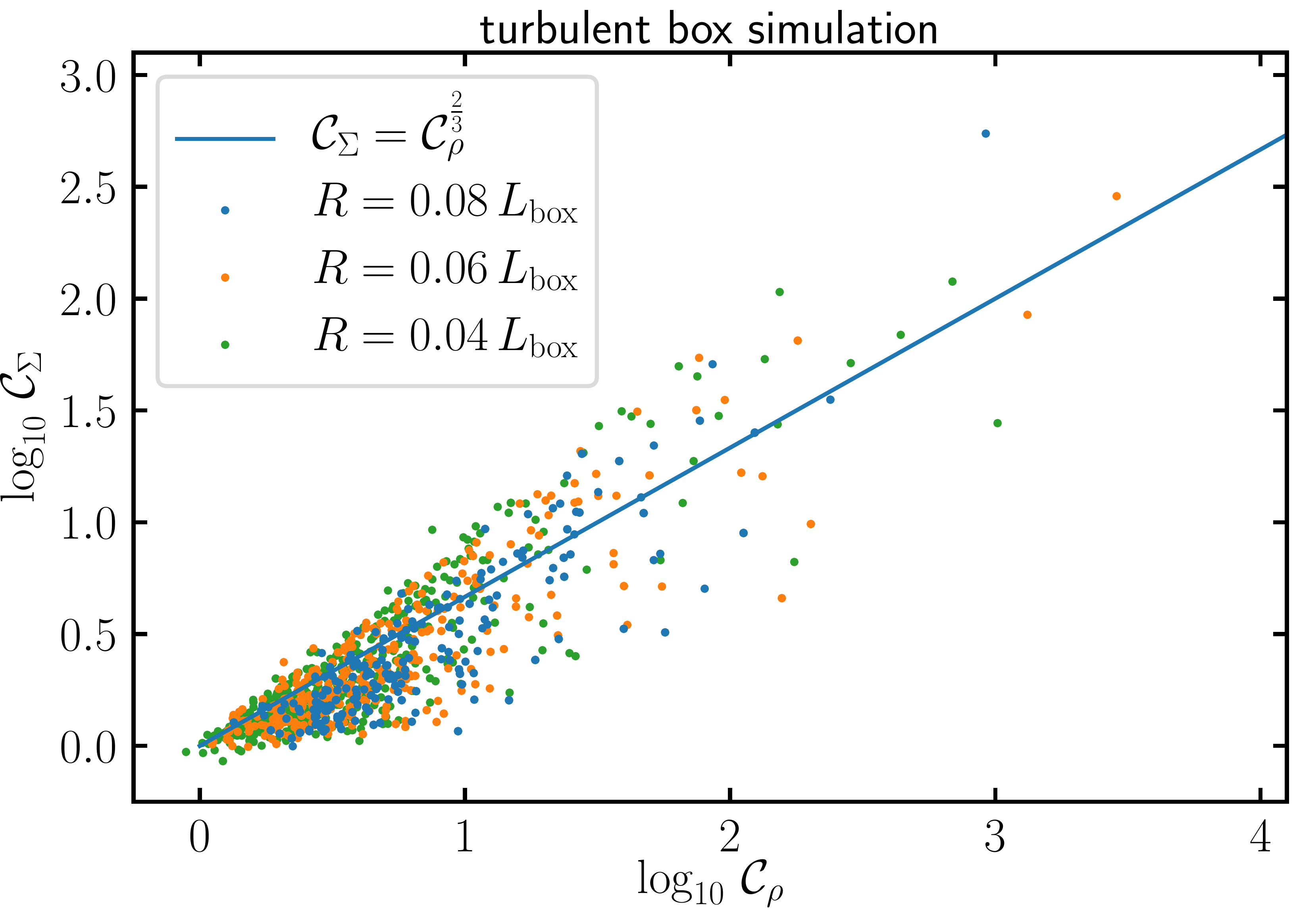}
  \includegraphics[width=.975\columnwidth]{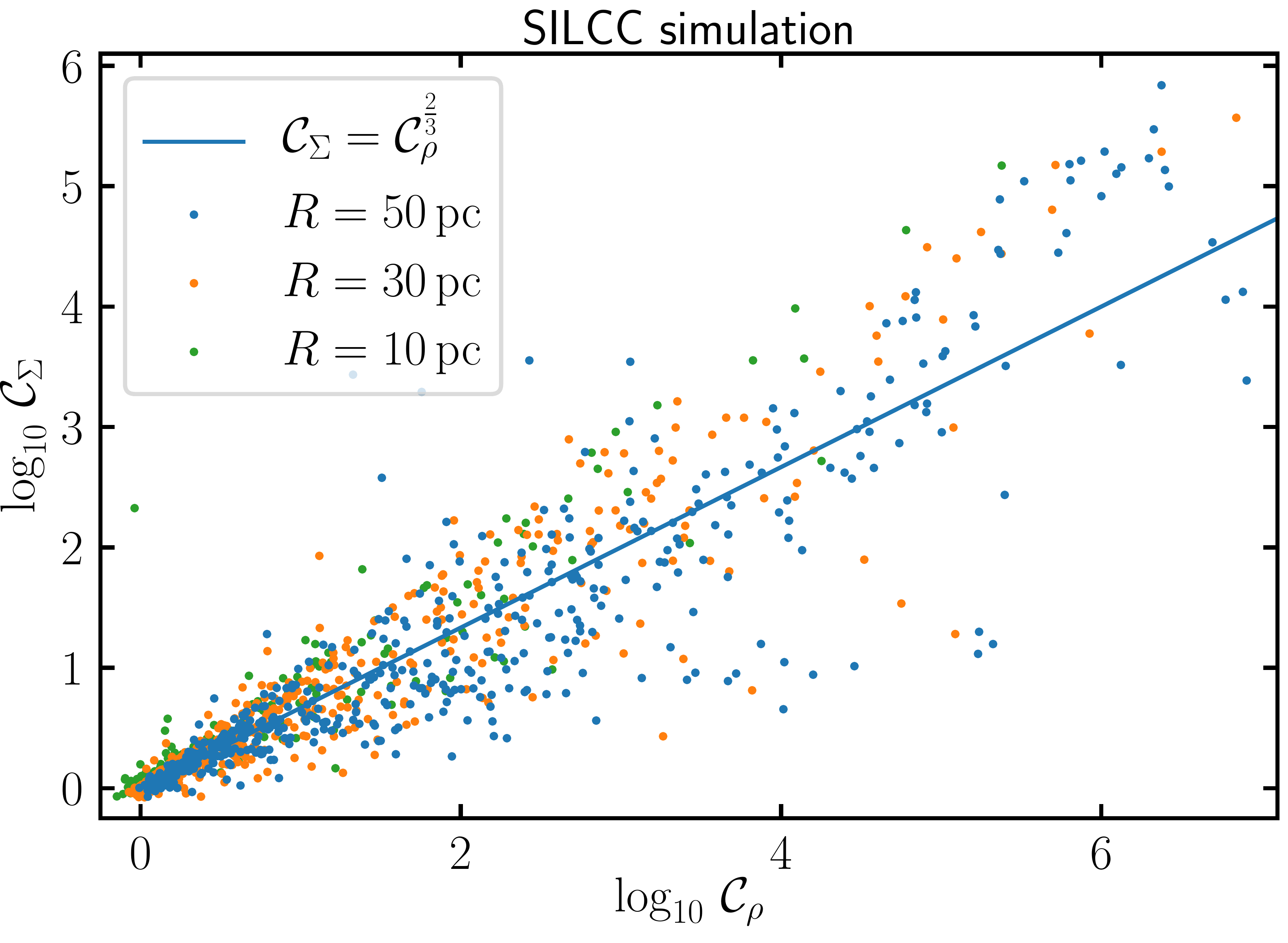}
  \oscaption{c_vs_c_all}{Three-dimensional clumping factor, $\C_\rho$ vs. projected clumping factor, $\C_\Sigma$. Each dot shows the clumping factors calculated for an individual spherical volume/surface of radius $R$ (different colors show different radii) extracted from the simulation domain of a turbulent box (left) and a patch of a galactic disc from the SILCC project (right, see text for more details). The projected clumping factor follows from the three-dimensional clumping factor as $\C_\Sigma=C_\rho^{2/3}$.}
  \label{fig:c_vs_c}
\end{figure*}

Equation~\eqref{eq:C_rho_vs_C_Sigma} establishes a connection between the two-dimensional column density clumping factor on the sphere surface, $\C_\Sigma$, and the three dimensional volume density clumping factor, $\C_\rho$, assuming a lognormal density PDF. In Fig.~\ref{fig:c_vs_c} we show the empirical relation between $\C_\rho$ and $\C_\Sigma$ as quantified from a turbulent box simulation and a high-resolution hydrodynamical simulation of a sizeable patch of a galactic disc from the SILCC simulation project \citep[][see Section~\ref{sec:sim} for more details]{GirichidisEtAl2018b}. Each dot in this figure represents an individual spherical region of radius $R$ extracted from the simulation domain within which we independently measure $\C_{\rho}$ and $\C_\Sigma$ from projecting the density inside the sphere onto its surface (see Section~\ref{sec:data} for more details).

We find that Eq.~\eqref{eq:C_rho_vs_C_Sigma} is fulfilled on average independent of the size of the sphere. However, we find substantial scatter that is growing as a function $\C_\rho$. Especially for the turbulent box result shown in the left panel we find that for low clumping factors $\C_\rho$ many spheres fall below the theoretical line of Eq.~\eqref{eq:C_rho_vs_C_Sigma}. We attribute this to the same physical effects that lead to a deviation of the density PDF from a purely lognormal behaviour (see Section~\ref{sec:turb}). As we have seen from Fig.~\ref{fig:c_vs_rho}, small clumping factors are preferentially realised in low- and high density environments which deviate most from a lognormal density PDF due to intermittency and larger-scale density correlations.

\label{lastpage}

\end{document}